\documentclass[iop, numberedappendix, appendixfloats]{emulateapj}
 \newcommand{\alfven}{Alfv\'{e}n}
 \newcommand{\alfvenic}{Alfv\'{e}nic}
 \newcommand{\oratio}{\mbox{O$^{7+}$/O$^{6+}$}}
 \newcommand{\pref}{\protect\ref}
 \newcommand{\hinode}{{\em Hinode}}

\begin{document}
 
\shorttitle{The Roots of the Fast Solar Wind}
\shortauthors{McIntosh, Leamon \& De Pontieu}
 \title{The Spectroscopic Footprint of the Fast Solar Wind}
\author{Scott W. McIntosh\altaffilmark{1}, Robert J. Leamon\altaffilmark{2,3}, Bart De Pontieu\altaffilmark{4}}
\email{mscott@ucar.edu, robert.j.leamon@nasa.gov, bdp@lmsal.com}
\altaffiltext{1}{High Altitude Observatory, National Center for Atmospheric Research, P.O. Box 3000, Boulder, CO 80307}
\altaffiltext{2}{ADNET Systems Inc., NASA Goddard Space Flight Center, Code 671.1, Greenbelt, MD 20771}
\altaffiltext{3}{Now at Department of Physics, Montana State University, Bozeman, MT 59717}
\altaffiltext{4}{Lockheed Martin Solar and Astrophysics Lab, 3251 Hanover St., Org. ADBS, Bldg. 252, Palo Alto, CA 94304}
 
\begin{abstract}
We analyze a large, complex equatorial coronal hole (ECH) and its immediate surroundings with a focus on the roots of the fast solar wind. We start by demonstrating that our ECH is indeed a source of the fast solar wind at 1~AU by examining in situ plasma measurements in conjunction with recently developed measures of magnetic conditions of the photosphere, inner heliosphere and the mapping of the solar wind source region. We focus the bulk of our analysis on interpreting the thermal and spatial dependence of the non-thermal line widths in the ECH as measured by {\em SOHO}/SUMER by placing the measurements in context with recent studies of ubiquitous \alfven{} waves in the solar atmosphere and line profile asymmetries (indicative of episodic heating and mass loading of the coronal plasma) that originate in the strong, unipolar magnetic flux concentrations that comprise the supergranular network. The results presented in this paper are consistent with a picture where a significant portion of the energy responsible for the transport of heated mass into the fast solar wind is provided by episodically occurring small-scale events (likely driven by magnetic reconnection) in the upper chromosphere and transition region of the strong magnetic flux regions that comprise the supergranular network.
\end{abstract}
 
\keywords{Sun: solar wind \-- Sun:magnetic fields \-- Sun:transition region \-- Sun:corona}
 
\section{Introduction}
Equatorial coronal holes (ECH) present spectroscopists with an unparalleled opportunity to study the source of the fast solar wind. These regions, while being shorter-lived, present us with a significantly less-complex line-of-sight than their well studied, polar cousins \citep[e.g.,][]{Dammasch+others1999,Hassler+others1999,Tu+others2005a, Xia2003, Xia2004, Popescu2004, Tian2009b}. In this Paper we revisit an ECH observed by the SUMER instrument \citep[][]{Wilhelm1995} on board the {\em Solar and Heliospheric Observatory} \citep[{\em SOHO};][]{Fleck1995} that was the primary subject of \citet{McIntosh2006, McIntosh2007}, while also providing support to the results of \cite{DePontieu2009}.
 
\citet{McIntosh2006} used \ion{Ne}{8} 770~\AA{} emission line spectroheliograms formed in the upper transition / low-corona plasma to demonstrate the magnetic complexity of the ECH and its surroundings. They noticed that the regions of strong outflow (blue Doppler shifts) are overlying photospheric regions of largest magnetic unipolarity that have an open coronal-scale magnetic field above. \citet{McIntosh2007} expanded on the initial work, investigating the relationship between the magnetic flux distribution in the photosphere and the transition region Doppler patterning on the supergranular scale. Using simultaneously observed spectroheliograms in the \ion{C}{4} 1548~\AA{} emission line they built a picture of a truly dynamic energy release into the plasma driven by the relentless evolution of the photospheric magnetic field, or ``Magnetic Carpet'' \citep[e.g.,][]{Schrijver1997} and placed the ``footprint'' of the outflow from the ECH in the middle of the transition region.
 
In this Paper we study a spectral feature of this ECH that has been ignored in our previous analyses of this region; the dependence on temperature of the non-thermal line widths of the transition region spectral lines in and around the coronal hole. To facilitate this investigation we will exploit the recent detection of chromospheric and coronal Alfv\'{e}n waves \citep[][]{DePontieu2007,Tomczyk2007}, and the relationship of the former with dynamic spicules \citep[e.g.,][]{DePontieu2007b}. The energetic and mass-loading signature of these heating events at low heights can be extended to the transition region \citep[e.g.,][]{McIntosh2008}, the active corona \citep[][]{DePontieu2009}, the quiet corona \citep[][]{McIntosh2009b}, and to a staple of coronal hole morphology - polar plumes \citep[][]{McIntosh2010b}. The diagnostic information provided by the non-thermal line widths provides further information about the footprint of the region which, as we will see, maps to a very high flow speed region in the solar wind as observed by the {\em Advanced Composition Explorer} \citep[ACE;][]{Stone1998}.
 
In Section~\pref{sanal2} we present some contextual observations of the region around the ECH and establish the physical connectivity of the region to the solar wind observed at 1AU following. Following that, in Sect.~\pref{obs}, we present the multi-thermal evolution of the non-thermal line widths in the ECH through the transition region and assess the impact of the measured line profile asymmetries on the line widths measured. Finally, in Sect.~\pref{results}, we discuss the data analysis  and results presented that offer us a look at the footprint of fast solar wind.

\begin{figure}
\epsscale{1.15}
\plotone{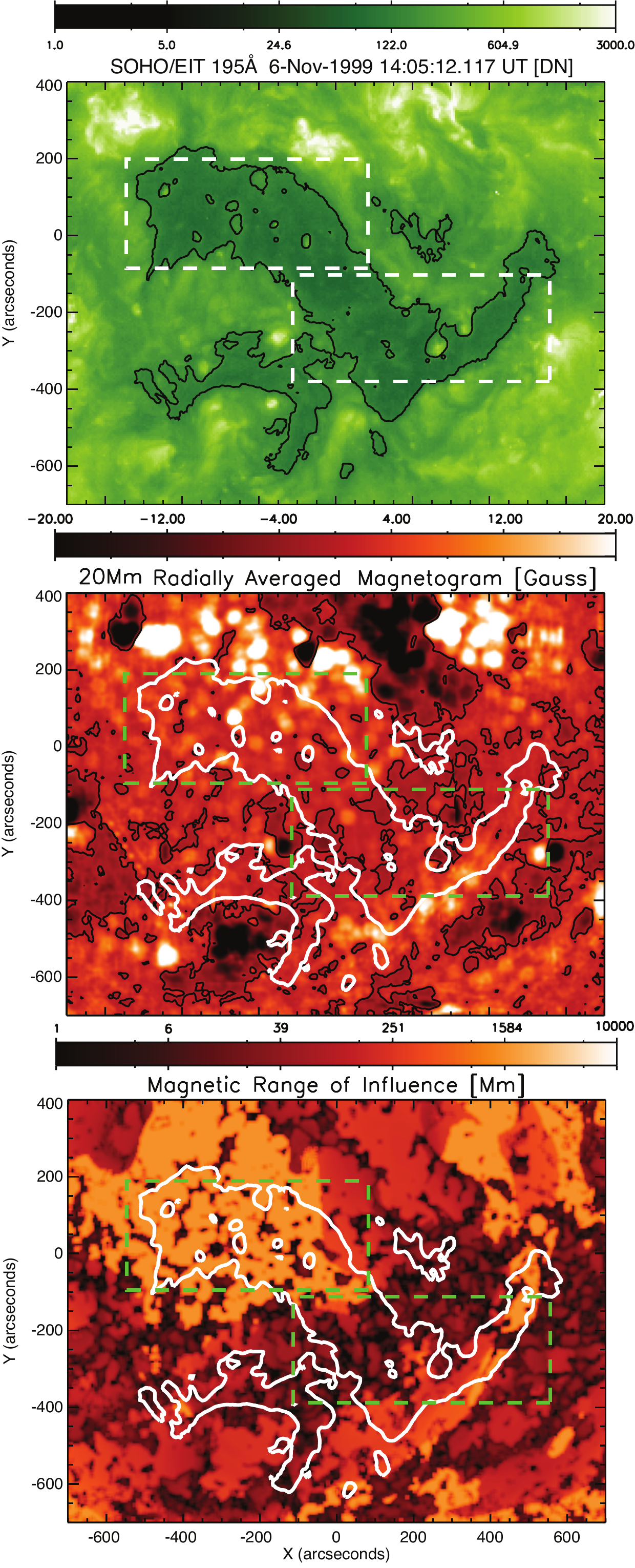}
\caption{Contextual information about the ECH region provided by the 195~\AA{} passband image of EIT (top), the supergranular-scale averaged LOS magnetogram (middle) and Magnetic Range of Influence (bottom) maps. In each case we show the 150~DN level of the 195~\AA{} as a proxy of the ECH boundary and the rectangular regions sampled by SUMER.\label{f1}}
\end{figure} 

\section{Is this ECH a Fast Wind Source at 1AU?}\label{sanal2}
 
The SUMER spectroheliograms of the ECH on 1999 November 6 and 7, as demonstrated in \citet{McIntosh2006} and Sect.~5 of \citet{McIntosh2007}, allowed us to begin a study of the coronal hole at disk center, minimizing interpretative error of both the line-of-sight spectral properties and magnetograms. Supporting observations of the magnetic environment of the ECH region are provided by abstractions of the SOHO/MDI \citep[][]{Scherrer1995} LOS magnetic field: the supergranular flux balance and Magnetic Range of Influence (MRoI) as defined, developed, and discussed in \citet{McIntosh2006}. Figure~\pref{f1} provides context of the extreme-UV and magnetic environment of the ECH from the Extreme-ultraviolet Imaging Telescope \citep[EIT;][]{Boudine1995} and MDI respectively. In each panel the ECH is outlined by a large closed contour determined simply from the 150~Data Number (DN) level of the 195~\AA{} EIT image\footnote{\citet{Habbal2008} explore this arbitrarily determined boundary in more detail removing the ``U-shaped'' filament channel region to the South and East of the ECH, see Sect.~\pref{results} for further discussion.} and the rectangular regions enclose the regions sampled by the individual SUMER spectroheliograms. We should reiterate that the supergranular flux balance map (middle panel) provides a measure of polarity ``dominance'' on a spatial scale of 20 Mm while the MRoI map (bottom panel) allows us to estimate the distance that separates partners in closed magnetic flux concentrations.
 
\begin{figure}
\epsscale{1.15}
\plotone{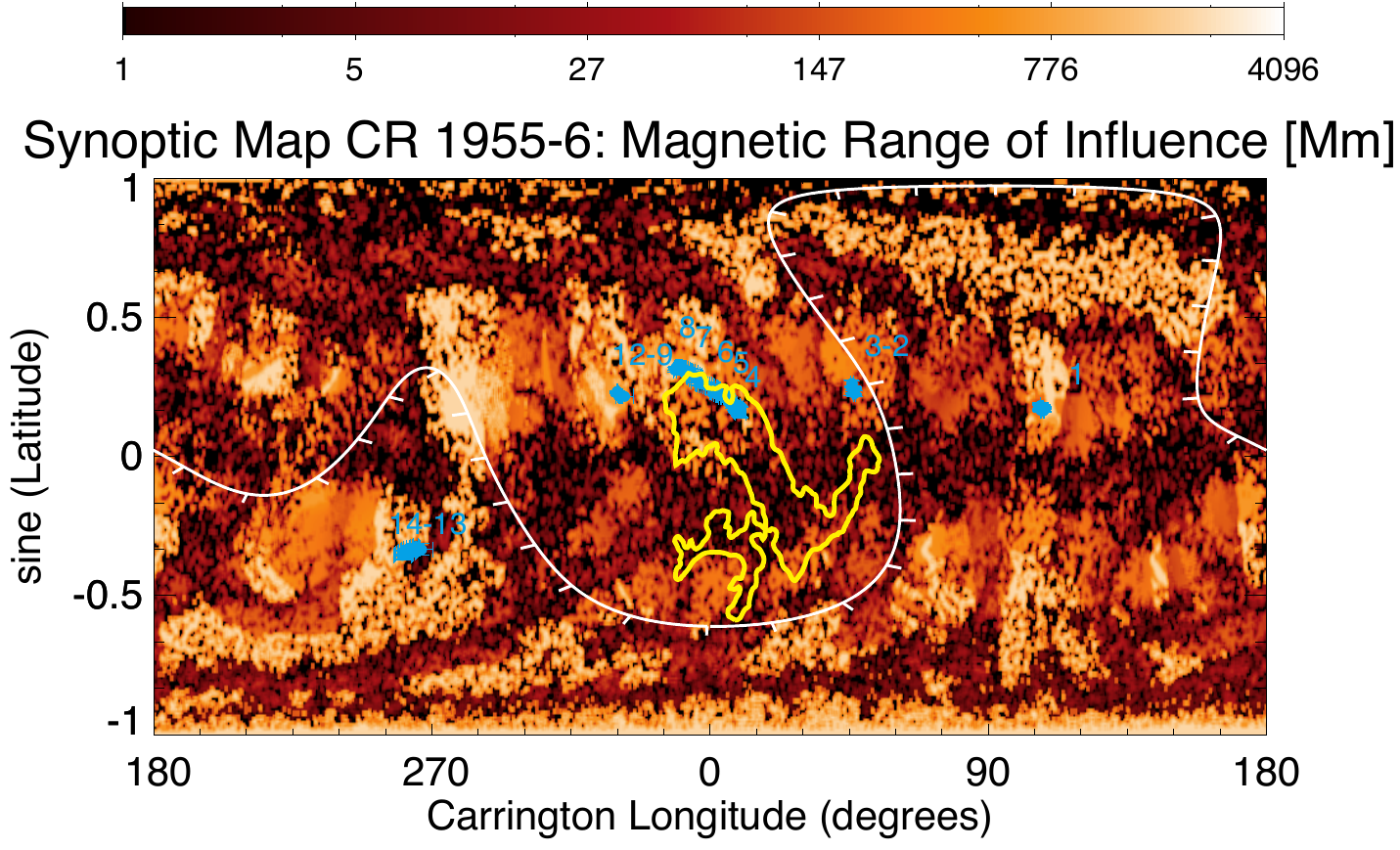}
\plotone{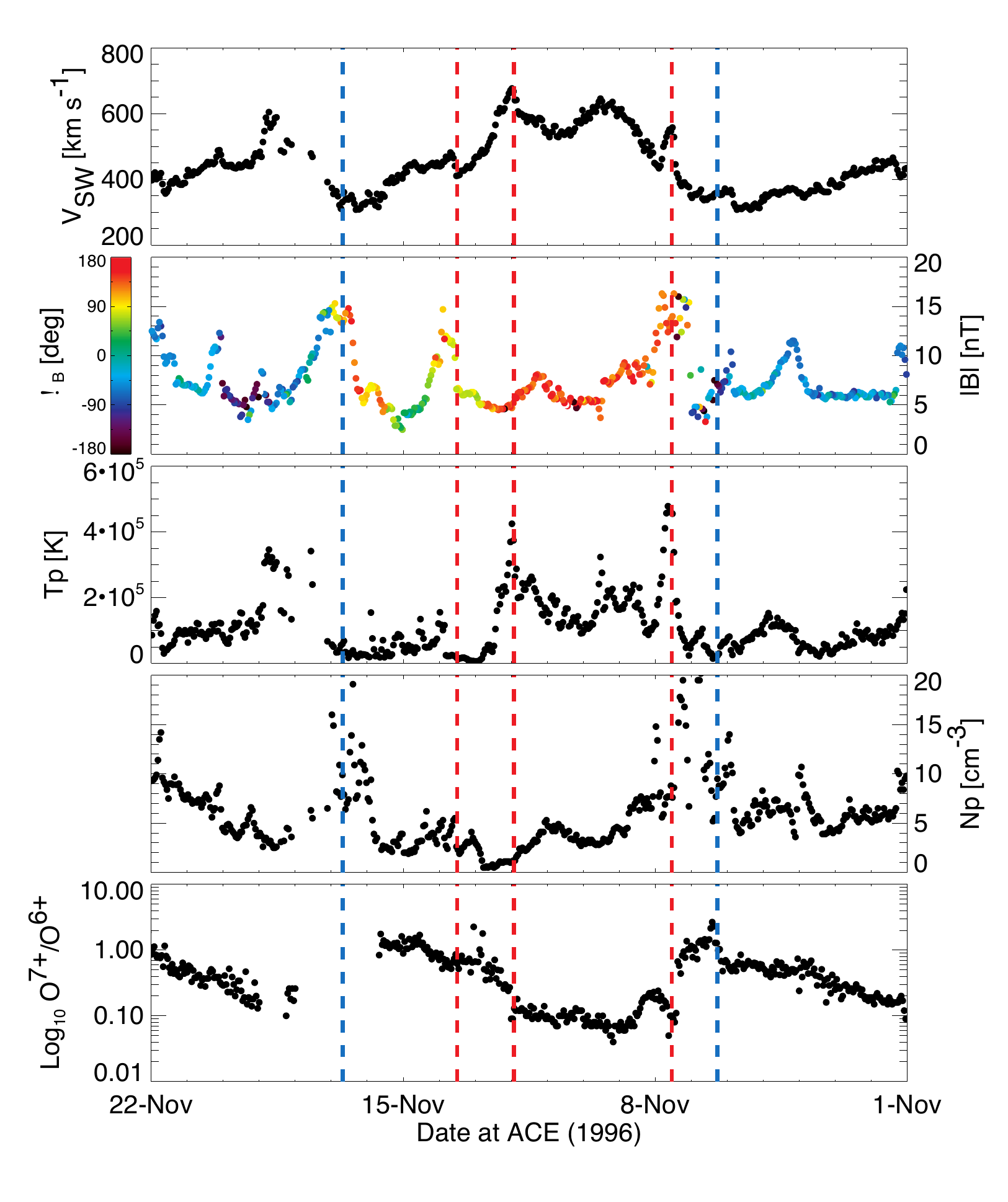}
\caption{Upper panel: Synoptic map of MRoI \protect\cite[]{Leamon2009} for CRs~1955-6, centered on Carrington Longitude~0\degr/360\degr. Yellow contour corresponds to the Coronal Hole boundaries defined in EIT 195~\AA\ counts by \protect\cite{McIntosh2006}. Blue dots show the track of the footpoint of the sub-terrestrial field line (determined from PFSS extrapolations), with error bars determined from the standard deviation of source surface ellipse footpoint separation (see text). Lower panels: Time series of solar wind speed, magnetic field strength and azimuth, proton temperature and density, and oxygen charge state ratio \oratio{} observed in situ by ACE\@. Note that time runs from right to left in these panels, as in the synoptic map. \label{f4}}
\end{figure}
 
However, in our previous papers we made no attempt to establish that this ECH maps to a high speed flow region of the solar wind at 1~AU. \citet{Leamon2009} noticed in their recent analysis that the footpoint of the sub-terrestrial field line, determined from Potential Field Source Surface extrapolations \citep[e.g.,][]{SchattenEA69}, jumps from patch to patch of locally high MRoI, and walks smoothly within each patch (e.g., Fig~\pref{f1} in that paper). The upper panel of Fig.~\ref{f4} shows a map of MRoI based on a merged synoptic magnetogram (downsampled by a factor of four---from $3600 \times 1080$~pixels to $900 \times 270$~pixels) comprising the trailing half of Carrington Rotation~1955 and the leading half of Carrington Rotation~1956.

In yellow, we show the same proxy for the ECH boundary: the smoothed 150~DN level of EIT 195~\AA{}. The blue dots track the sub-terrestrial field line footpoints for the first two weeks of November 1999 (one dot per 96-min MDI magnetogram). Repeating the analysis of \cite{Leamon2007,Leamon2009}, we trace 32 additional field lines (arranged on the perimeter of an ellipse on the source surface with semimajor axes $5\degr \times 2.5\degr$) back to the photosphere. The additional field lines allow us to compute the standard deviation of the separation between their footpoints and that of the sub-terrestrial field line - providing error estimates for the location of each sub-terrestrial footpoint. To indicate the progression of time in the plot we label the magnetogram closest to noon each day.

We see that the footpoint does not progress smoothly across the map---rather, it jumps from one point to the next. It suddenly jumps (by $32\degr$ of longitude) into the equatorial coronal hole after spending much of the first two days close to AR~8742. A similarly large jump is seen on the November 13 after the footpoint has spent much of November 9-12 close to AR~8760.

The lower panels of Fig.~\ref{f4} shows in situ data observed at 1~AU by ACE, from top to bottom: the solar wind speed; magnetic field strength and (color-coded) azimuth \cite[MAG;][]{SmithEA98_ACE}; proton temperature \&\ density, and Oxygen charge state ratio \oratio{} \cite[SWIMS;][]{GloecklerEA98_short}. The time shown in these panels is that {\em at ACE}, and runs from right to left, as in the synoptic magnetogram above them. Notice the five vertical dashed lines that correspond to the time of the first magnetogram after five significant jumps: blue for the South-North crossing of the heliospheric current sheet (1999 November 2, 12:51UT); red for when the footpoint first enters the coronal hole (1999 November 4, 14:23UT), leaves the EIT-intensity defined boundary (1999 November 8, 03:12UT), and the jump from the coronal hole to AR~8760 (1999 November 9, 12:47UT); and again in blue the second crossing of the heliospheric current sheet (1999 November 13, 04:47UT). We note that while the first current-sheet crossing is muddied by the major change in wind parameters associated with fast coronal hole outflow, the second shows the classic signature of a current-sheet crossing: as the magnitude increases (i.e., field-line packing) and the field direction rotates smoothly through over $180\degr$. 

The propagation delays from the Sun to 1~AU are calculated from the solar wind forecast model of \cite{Leamon2007}. The predictions for the current sheet crossings are both about half a day too late based on field and plasma signatures, but correlate well with a small increase in \oratio{} (unfortunately, the second current sheet crossing occurs during a gap in SWICS coverage). However, the jump into the coronal hole manifests itself with low \oratio\ conditions (bottom panel) at almost precisely the right time. Simultaneously, we see an interplanetary shock, a step in wind speed, and associated spikes in the proton density and temperature, as the shock compresses and heats the plasma, and increases the field strength. The low \oratio\ conditions typical of a coronal hole outflow are bracketed by the first two red lines. There is a small shock and a discontinuity in field strength at the third red dashed line that is associated with the footpoint's jump to AR~8760 where it remains until the second current sheet crossing. In summary, the physical connectivity of the ECH to the solar wind observed at 1~AU is clear.

\section{SOHO/SUMER Observations, Analysis, and Discussion}\label{obs}
\begin{figure*} 
\epsscale{1.15}
\plotone{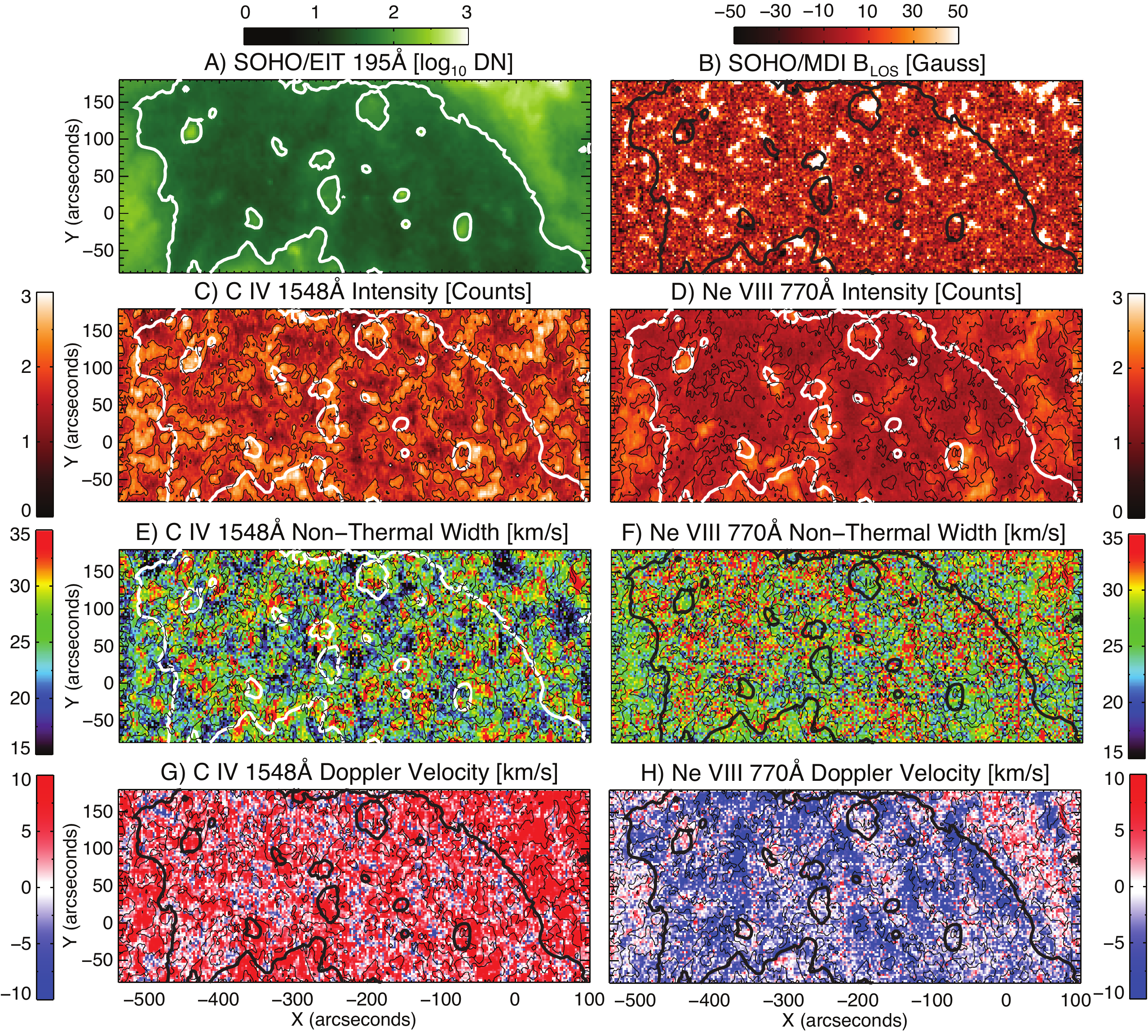}
\caption{Top row: the EIT 195~\AA{} image and composite MDI LOS magnetogram (see text) for the SUMER spectroheliogram of November 6 1999. Subsequent rows show the fitted line intensities, non-thermal line widths and absolutely calibrated Doppler velocities of the \ion{C}{4} 1548~\AA{} (left) and \ion{Ne}{8} 770.4~\AA{} (right) emission lines. In each case the fitted spectral information (panels C-H) are overlaid with the 150 DN EIT 195\AA{} intensity contour as the ECH boundary (thick line) and the 150 count contour of \ion{C}{4} intensity (thin black line) as a proxy for the supergranular network boundary. \label{f2}} 
\end{figure*}

\begin{figure*}
\epsscale{1.}
\plotone{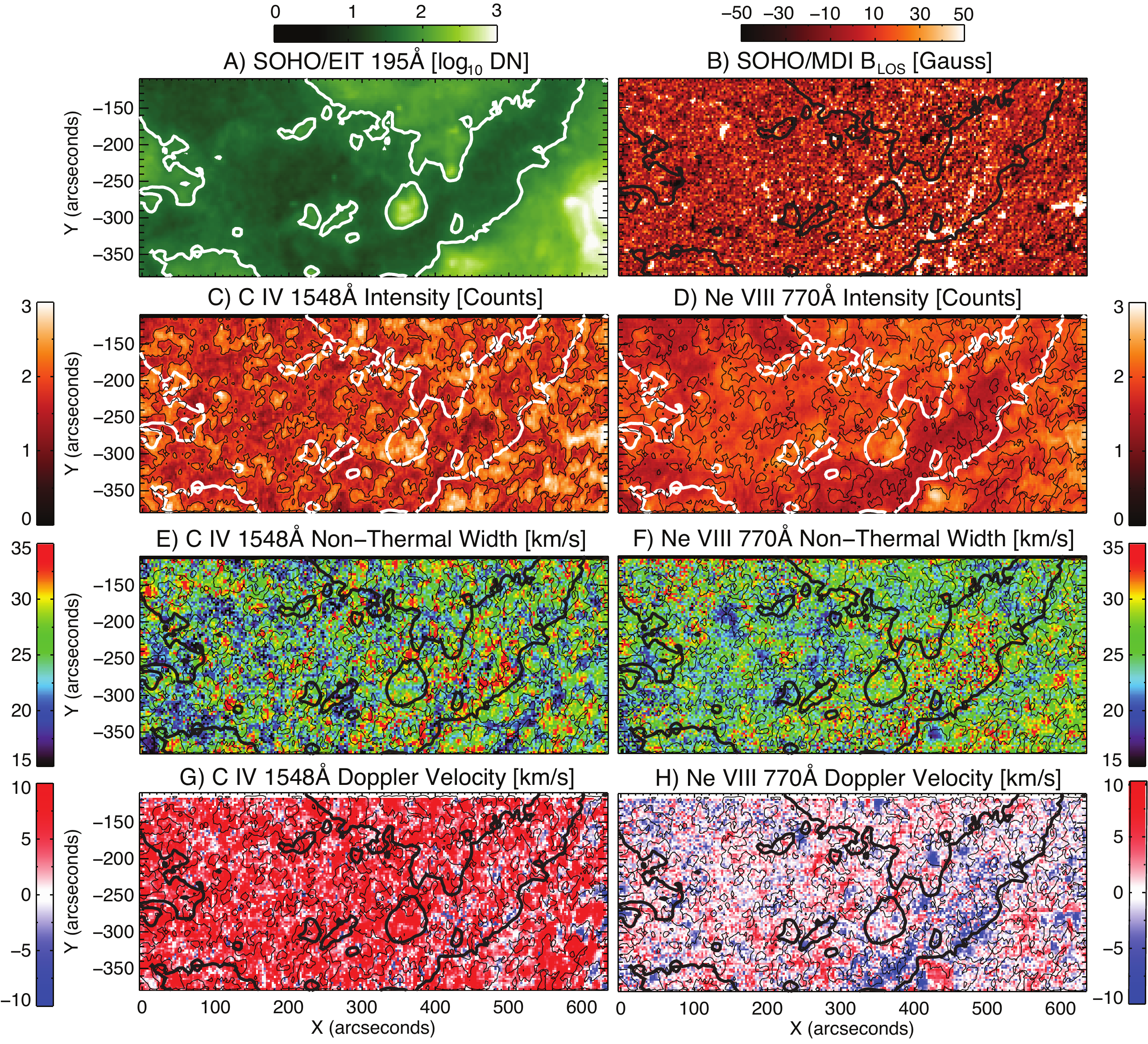}
\caption{Top row: the EIT 195~\AA{} image and composite MDI LOS magnetogram (see text) for the SUMER spectroheliogram of November 7 1999. Subsequent rows show the fitted line intensities, non-thermal line widths and absolutely calibrated Doppler velocities of the \ion{C}{4} 1548~\AA{} (left) and \ion{Ne}{8} 770.4~\AA{} (right) emission lines. In each case the fitted spectral information (panels C-H) are overlaid with the 150 DN EIT 195\AA{} intensity contour as the ECH boundary (thick line) and the 150 count contour of \ion{C}{4} intensity (thin black line) as a proxy for the supergranular network boundary. \label{f3}} 
\end{figure*}

The spectroheliograms studied in \citet{McIntosh2006, McIntosh2007} were taken as part of a SUMER campaign to study the evolution of an ECH as it transited the disk from 1999 November 3-8 - each has 3\arcsec{} stepping and an exposure time of 150s, with each ``pixel'' representing the cumulative spectra of nine individual 1\arcsec{} SUMER spectra (in a 3\arcsec{}$\times$3\arcsec{} array), that is needed to boost the signal-to-noise of the dataset (see also Appendix~\pref{sec:blends}). For each observing sequence in the campaign, a spectroheliogram is formed in three emission lines that span the upper chromosphere, transition region, and low-corona (\ion{Si}{2} 1534~\AA{} \-- \ion{C}{4} 1548~\AA{} \-- \ion{Ne}{8} 770~\AA{} in second spectral order\footnote{These emission lines are formed (in equilibrium) at  30,000, 100,000 and 600,000K \citep[][]{Mazzotta1998}.}) which have been calibrated (absolutely) in wavelength using the technique of \citet{Davey2006}. The first (coolest) of these lines is omitted from the present analysis due to the presence of a strong \ion{Si}{1} blend in the red wing of the line, and reduced signal-to-noise\footnote{The potential impact of \ion{Si}{1} blends on the \ion{C}{4} and \ion{Ne}{8} lines is discussed briefly in the body of the Paper, and in more detail in Appendix A.}.

Figures~\pref{f2} and~\pref{f3} provide the detailed spectral context for the two SUMER spectroheliograms (1999 November 6 14:00-00:00UT and 1999 November 7 (00:35-10:16UT)) that span the equatorial coronal hole studied (the rectangular regions outlined in Fig.~\pref{f1}). In each case we show the EIT 195~\AA{} image and a composite MDI line-of-sight (LOS) magnetogram\footnote{The magnetogram shown is a composite of the four 96-minute cadence full-disk magnetograms taken over the duration of the SUMER spectroheliogram. The composite is needed to take into account the considerable evolution of the photospheric magnetic field that has taken place over the 7-plus hours of SUMER observation.}, the line intensities, non-thermal line widths \citep[$v_{nt}$; determined from the recipe in][and below]{McIntosh2008} and absolutely calibrated Doppler velocities of the \ion{C}{4} 1548~\AA{} and \ion{Ne}{8} 770.4~\AA{} lines in the left and right columns respectively. In each case the spectral information is overlaid with the 150 count intensity contour of the \ion{C}{4} intensity as a proxy for the supergranular network vertices. We note that the visibility of the network vertex pattern in \ion{Ne}{8} (panel D) is NOT immediately obvious in either example \citep[a point discussed in other papers pertaining to this line, e.g.,][]{Xia2003}. 

Comparing the composite magnetograms and maps of $v_{nt}$ we see a general correspondence between $v_{nt}$ and the LOS field strength in both emission lines. In both figures, for \ion{C}{4}, we see a mix of enhanced $v_{nt}$ in the supergranular vertices where the strongest regions of magnetic flux are located and in regions straddling the intensity-defined network vertices (see, e.g., the region around 475\arcsec, -275\arcsec{} in panel E of Fig.~\pref{f3}). The periphery of the network vertices is exactly where we would reasonably expect to have regions of highly inclined magnetic field. Further, we notice a strong visual correspondence between the regions of significant \ion{Ne}{8} blue Doppler shift and enhanced $v_{nt}$ that are not only isolated to the supergranular network vertices. The values of $v_{nt}$ in the region of \ion{Ne}{8} enhanced emission in the center of panel F in Fig.~\pref{f3}, and particular those in the network vertices are reduced ($\sim$20km/s) compared to the regions of strong outflow (30-35km/s). Based on the analysis of \citet{McIntosh2007} this central portion of enhanced \ion{Ne}{8} emission is, to all intents and purposes, quiet sun, not magnetically open and should be different from the open regions of the field-of-view. Taking the two figures together, we see that the first and second moments (line position/shift, and width) of the \ion{Ne}{8} line are strongly influenced by the open coronal field and outflow of the ECH. Investigating the relationship between the magnetic environment on the global (and supergranular) scales, the $v_{nt}$ and profile symmetry in both spectral lines motivates the remainder of this Section.

\begin{figure}
\plotone{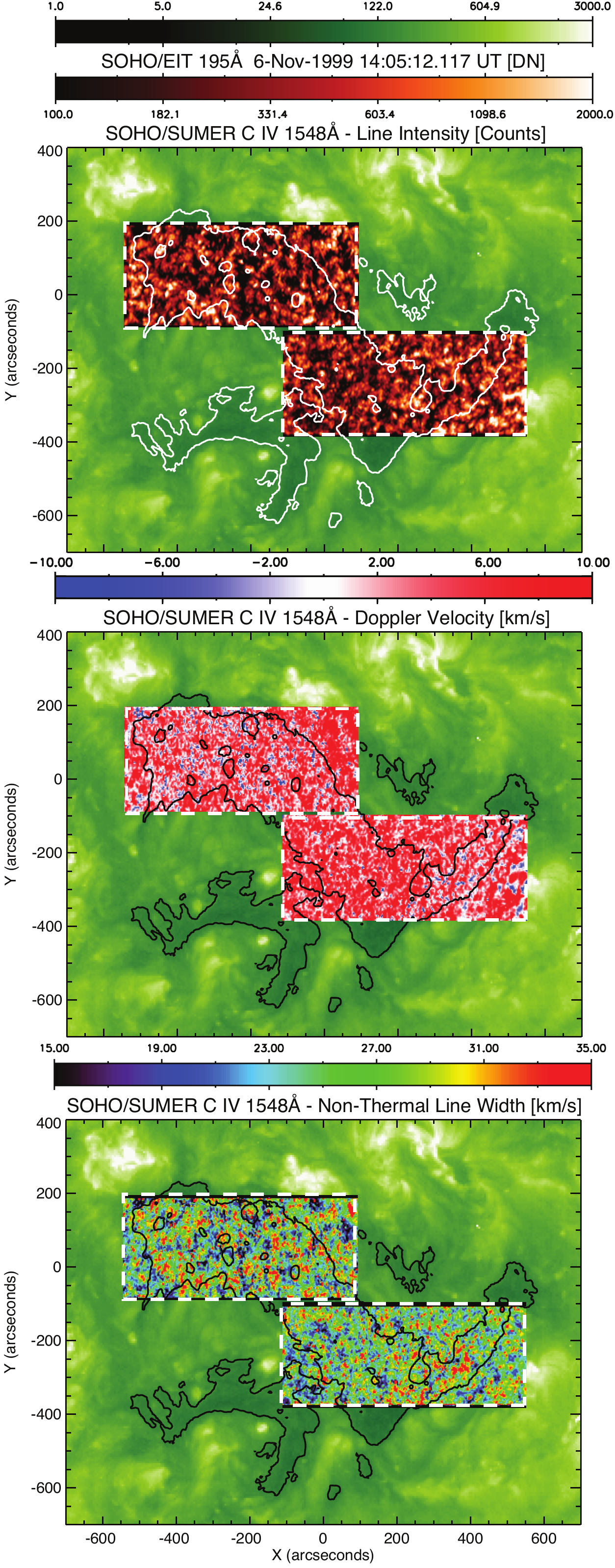}
\caption{Comparing the spatial behavior of the SUMER spectroheliograms in the \ion{C}{4} 1548~\AA{} emission line overlaid on the EUV image from Fig.~\pref{f1}. The top panel shows the peak intensity of the line, the middle panel shows the Doppler velocity in the region while the bottom panel shows the variation in non-thermal line width ($v_{nt}$). In each case the contours shown designate the EUV determination of the coronal hole boundary. \label{f5}}
\end{figure}

\begin{figure}
\plotone{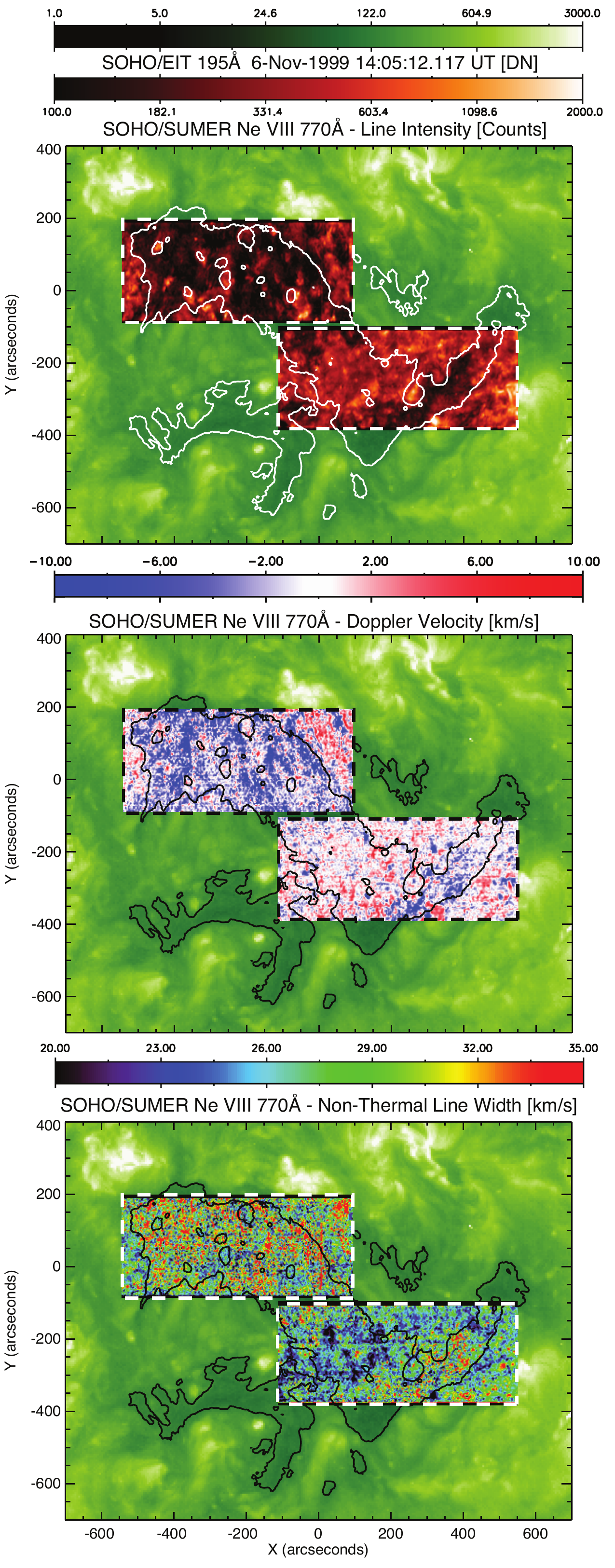}
\caption{Comparing the spatial behavior of the SUMER spectroheliograms in the \ion{Ne}{8} 770~\AA{} emission line overlaid on the EUV image from Fig.~\pref{f1}. The top panel shows the peak intensity of the line, the middle panel shows the Doppler velocity in the region while the bottom panel shows the variation in non-thermal line width ($v_{nt}$). In each case the contours shown designate the EUV determination of the coronal hole boundary. \label{f6}}
\end{figure}

\subsection{Non-Thermal Line Width Behavior With Temperature}\label{sanal1}
 
Non-thermal line widths are extracted from the emission line profiles of \ion{C}{4} and \ion{Ne}{8} using the method discussed in Sect.~2 of \citet{McIntosh2008}. Using a single Gaussian Genetic Algorithm fit to the emission line profiles in each case \citep[e.g.,][]{McIntosh1998} we compute the non-thermal line width $v_{nt}$ $(= w_{1/e}^2 - w_{inst}^2 - w_{th}^2$) of the line profile at 1/e of the peak intensity where $w_{1/e}$ is determined from the FWHM of the fitted Gaussian ($= 2\sqrt{\ln(2)}\;w_{1/e}$), $w_{th} (= \sqrt{2k_{B}T_{e}^{\ast} / m_{ion}}$ for an ion of mass $m_{ion}$ and peak formation temperature of $T_{e}^{\ast}$ \-- assuming that the ion and electron temperatures are equal) is the thermal width, and $w_{inst}$ is the SUMER instrumental width which is determined from the SUMER solarsoft routine {\tt CON\_WIDTH\_FUNCT\_3.PRO}. Note that for this part of the SUMER spectrum the velocity scale in each spectral pixel is of the order of 12~km/s/px and for the typical 25 pixel wide spectral window we can centroid the Gaussian profile and determine its 1/e width to within 0.1~pixels \citep[][]{McIntosh1998}.

Figure~\pref{f5} shows the variation of the line peak intensity (top), Doppler velocity (middle) and  $v_{nt}$ (bottom) in the two SUMER \ion{C}{4} 1548~\AA{} spectroheliograms studied\footnote{Compare with the analysis of \citet{Xia2003}.}. Comparing the top and middle panels we can clearly see the persistent transition region red-shift and its enhancement in the supergranular network \citep[see, e.g.,][and many others]{Gebbie1981} throughout the region. We also note the spatial variation of the Doppler patterning as there is a considerable increase in the area of the images showing blue-shifted \ion{C}{4} emission. Most obvious in the upper left and lower right regions of the field of view, this change in Doppler velocity ``partitioning'' was the subject of the discussion in Sect.~5 of \citet{McIntosh2007} and occurs in the regions of largest magnetic flux imbalance and MRoI. We interpreted the increased number of localized \ion{C}{4} blue-shifts to indicate the presence of small concentrations of outflowing material on the supergranular boundary interiors and believe that these were most likely macrospicules, driven by magnetic carpet forced reconnection. Comparing the top and bottom panels we see the strong relationship between the $v_{nt}$ and line intensity - again, the cell interiors show values around 20km/s which increase to $\sim$27km/s on the boundaries with some larger concentrations of enhanced $v_{nt}$ ($\sim$35km/s) in the upper left and lower right of the ECH region with noticeably less network/interior contrast in the intermediate (mixed polarity) region. 

 
\begin{figure*}
\plotone{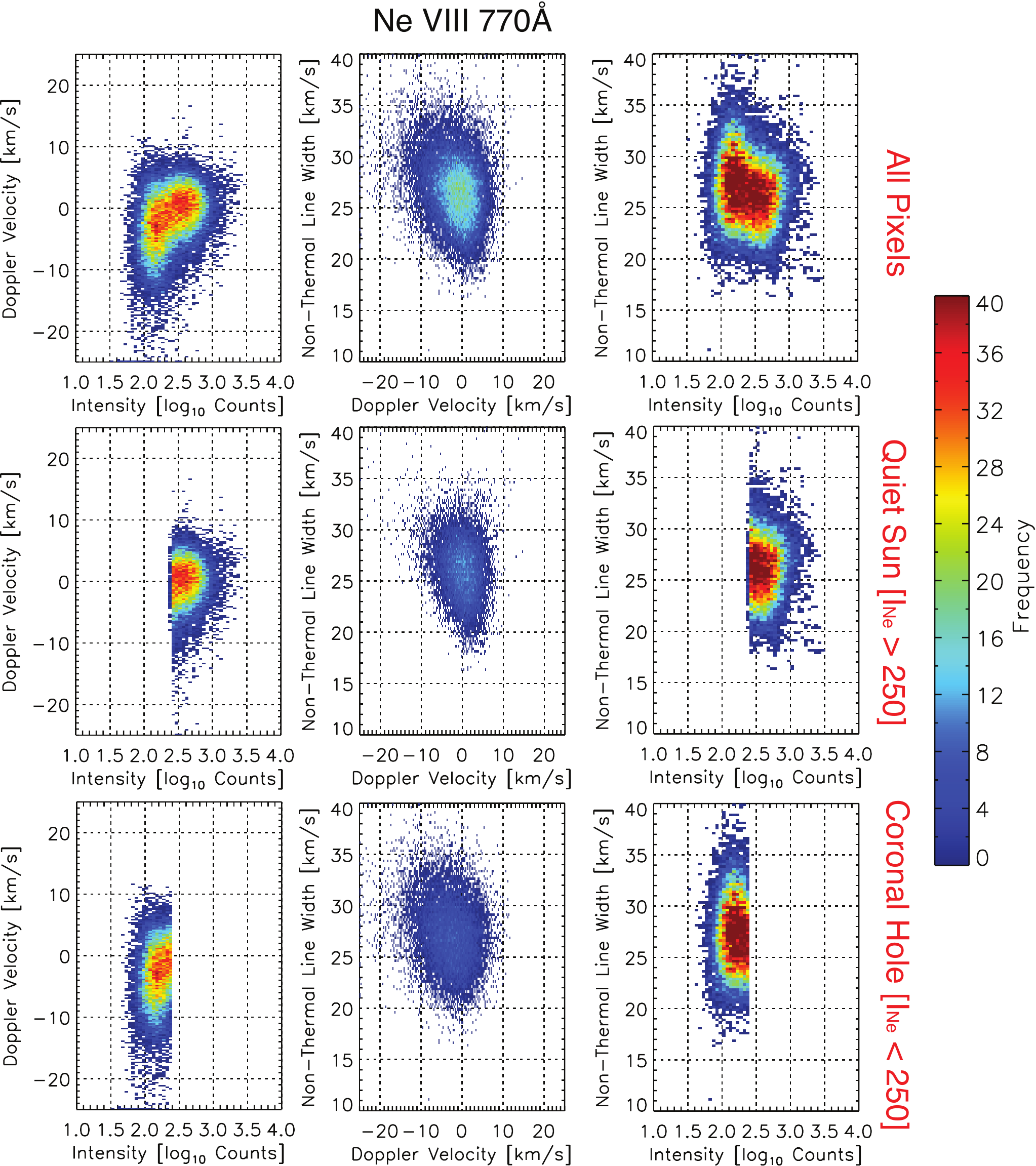}
\caption{Scatter plots of line Intensity, Doppler velocity and non-thermal line width in the \ion{Ne}{8} 770~\AA{} emission line from both SUMER spectroheliograms. The left column of panels show the relationship between the Doppler velocity and intensity, the central column shows the relationship between the non-thermal line width and Doppler velocity, the right column shows the non-thermal line width - intensity relationship. While the top row shows the distribution for all pixels, the two lower figures show the plots for the quiet sun (centre row) and coronal hole (bottom row) regions where the boundary between the two has been drawn at an \ion{Ne}{8} intensity of 250 counts. \label{f7}}
 \end{figure*}

\begin{figure}
\epsscale{1.15}
\plotone{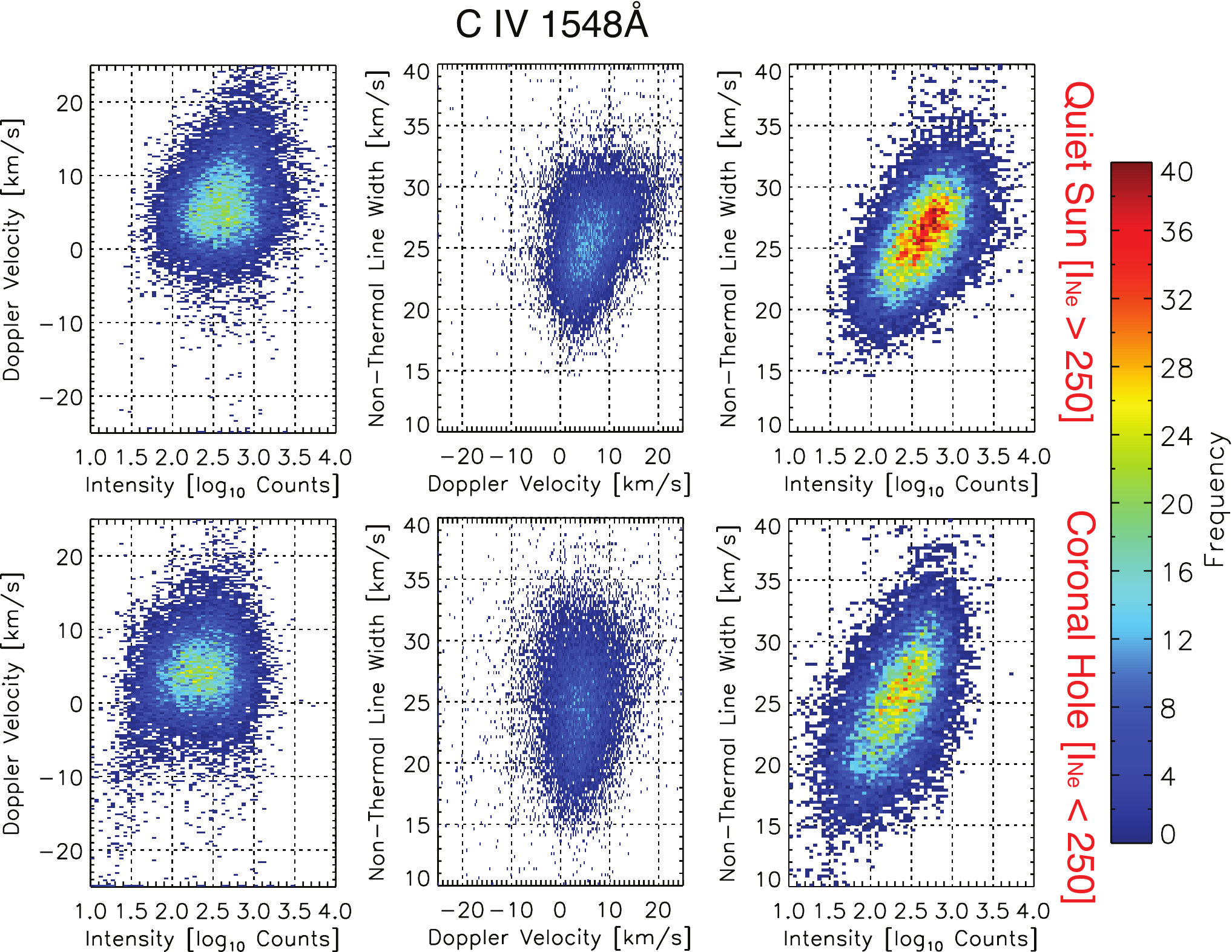}
\caption{Scatter plots of line intensity, Doppler velocity and non-thermal line width pairs in the \ion{C}{4} 1548~\AA{} emission line from both SUMER spectroheliograms in the quiet sun (top row) and coronal hole (bottom row) regions; where the boundary between the two is delineated at an \ion{Ne}{8} intensity of 250 counts. \label{f8}}
\end{figure} 
 
Figure~\pref{f6} shows the variation of the line peak intensity (top), Doppler velocity (middle) and $v_{nt}$ (bottom) in the two SUMER \ion{Ne}{8} 770~\AA{} spectroheliograms in a broader context, adding significantly to the versions published in \citet{McIntosh2006} and \citet{McIntosh2007}. Immediately, we see the correspondence between the regions of strong blue Doppler shift, enhanced $v_{nt}$ and reduced line intensities. In the ECH, the largest blue-shifts ($\sim$15km/s) have $v_{nt}$ of the order of 35km/s, but we see that the region (inside the EIT 195~\AA{} determined ECH boundary) where the \ion{Ne}{8} line intensity reaches quiet sun levels is exactly where we see a net reduction in the Doppler velocity ($\sim$0km/s) and $v_{nt}$ ($\sim$22km/s). This correspondence between the intensity and Doppler velocity was first noted in \citet{McIntosh2006} where the strength of the the blue-shift observed in \ion{Ne}{8} and the strength of photospheric magnetic imbalance and ``openness" were linked. In the middle region of the EUV ECH, the \ion{Ne}{8} emission is enhanced, the magnetic flux imbalance is low and the plasma appears to close locally (small values of the MRoI). This region also shows values of $v_{nt}$ that appear to be anti-correlated with the bright emission present - vastly different from those seen in the open regions of large magnetic imbalance. We note that the other regions of enhanced emission inside the EUV ECH show reduced Doppler velocity and $v_{nt}$, although the reduction is not as pronounced. The close correspondence in the spatial behavior of the \ion{Ne}{8} spectral diagnostics (the variation in intensity can be easily connected to Doppler velocity and $v_{nt}$) hints that the plasma at this temperature may be more simply structured than the plasma emitting the cooler \ion{C}{4} emission.

The discussion presented in \cite{McIntosh2006, McIntosh2007} and \cite{Habbal2008} focuses on the apparent discrepancy between the \ion{Ne}{8} definition of where the ECH is rooted (using spectral diagnostics) and where the EUV corona would have you believe that it is. As we have seen, the spectroscopic diagnostics observed in the \ion{C}{4} and \ion{Ne}{8} lines can differ significantly in the dark regions of the EUV corona versus the brighter regions -  suggesting, naturally, that the spectroscopic measurements provide a better indication of magnetic openness of the region than the apparent column depth of the EUV corona does. In order to quantify the differences in $v_{nt}$ imposed by changes in the global magnetic topology we construct scatter plots of line intensity, $v_{nt}$ and Doppler velocity. Unlike the work of \cite{McIntosh2007}, we will not attempt to rigidly differentiate between the spectroscopic diagnostics of the supergranular cell interior and boundaries using (subjective) intensity thresholding of \ion{Si}{2} - we are more interested in the impact of the open and closed magnetic topologies on the $v_{nt}$ values of the upper TR plasma. A further reason not to differentiate between cell boundaries and interiors here is dictated by the significantly reduced supergranular network boundary/interior contrast of \ion{Ne}{8} \citep[or in most lines formed above $\sim$600,000K see, e.g.,][]{Withbroe1983}. Further, by studying the continuity of the emission across the whole supergranule we will be able to investigate the subtle changes in intensity, velocity, and $v_{nt}$ contrast that are representative of the physical conditions of the boundary and interior together with the global scale variations in a single three-dimensional plot with reasonable confidence. 

First, we must consider the spectroscopic diagnostics of \ion{Ne}{8} as they provide the most obvious delineation of the open ECH plasma. Under the assumption that we can safely combine the statistics of both absolutely calibrated spectroheliograms we show the scatterplots of the line intensity, Doppler velocity, and $v_{nt}$ in Fig.~\pref{f7}. Notice the appearance of two distinct populations, or clusters of points, in the intensity-$v_{nt}$ and intensity-velocity panels that can be separated at an intensity level of 250~counts. We see that the lower intensity set of points have typically larger values of $v_{nt}$ and blue-shifted velocities while the mean Doppler velocity of the brighter pixels is $\sim$0km/s and the profiles are narrower than the higher intensity set. In the earlier papers \citep[][]{McIntosh2006,McIntosh2007} this last point was designated as the key difference between the ECH and quiet sun plasmas, adding the significantly enhanced $v_{nt}$ to the picture provides important information and consistency checks towards understanding the driver of the wind streams likely to originate from the region. From this point on we will adopt the 250~count intensity level of \ion{Ne}{8} as the ECH boundary for the transition region plasma. In the two lower rows of Fig.~\pref{f7} we show the decomposition of the top row distributions into quiet sun (I $>$ 250; center) and coronal hole (I $\le$ 250; bottom) components. If this intensity threshold marks the coronal hole boundary in \ion{Ne}{8} then it appears to form a spatially sharp boundary region in that the spectral quantities change rapidly across the intensity threshold.

The panels of Fig.~\pref{f8} show the scatter plots of the \ion{C}{4} line spectral diagnostics in the re-designated quiet sun (top) and coronal hole (bottom) regions. In the quiet sun we see the strong dependence of intensity and red Doppler shifts - characteristic of most transition region lines in the supergranular network \citep[see, e.g.,][]{Gebbie1981,Chae1998b,Peter1999}. We also see the strong relationship between the line intensity and $v_{nt}$ as was demonstrated for this line by \citet{Dere1984} although, unlike \citet{Dere1984}, we also see a relationship between plasma red-shifts and enhanced values of $v_{nt}$. 
The coronal hole scatter plot has a similar shape to that from the quiet sun region although, in general all of the distributions tend to shift to lower intensities and Doppler velocities. We again see a strong  $v_{nt}$ - intensity relationship that shows a steeper gradient than its quiet sun counterpart. We also see a steeper relationship between Doppler velocity and $v_{nt}$ which is, in part, due to the reduction of the mean network redshift in the coronal hole region - also discussed in \citet{McIntosh2007}. The steeper relationship between the intensity and $v_{nt}$ inside the coronal hole enforces our earlier (visual) impression that the network cells of the ECH show a higher interior/boundary contrast in $v_{nt}$. It appears also that the brightest coronal hole network vertices in \ion{C}{4} show enhancements in $v_{nt}$ of the order of a few km/s over their quiet sun counterparts. It remains, however, that the coronal hole and quiet sun distributions are largely similar; a detail that we will discuss in Sect.~\pref{sec:cons}.

\begin{figure*} 
\plotone{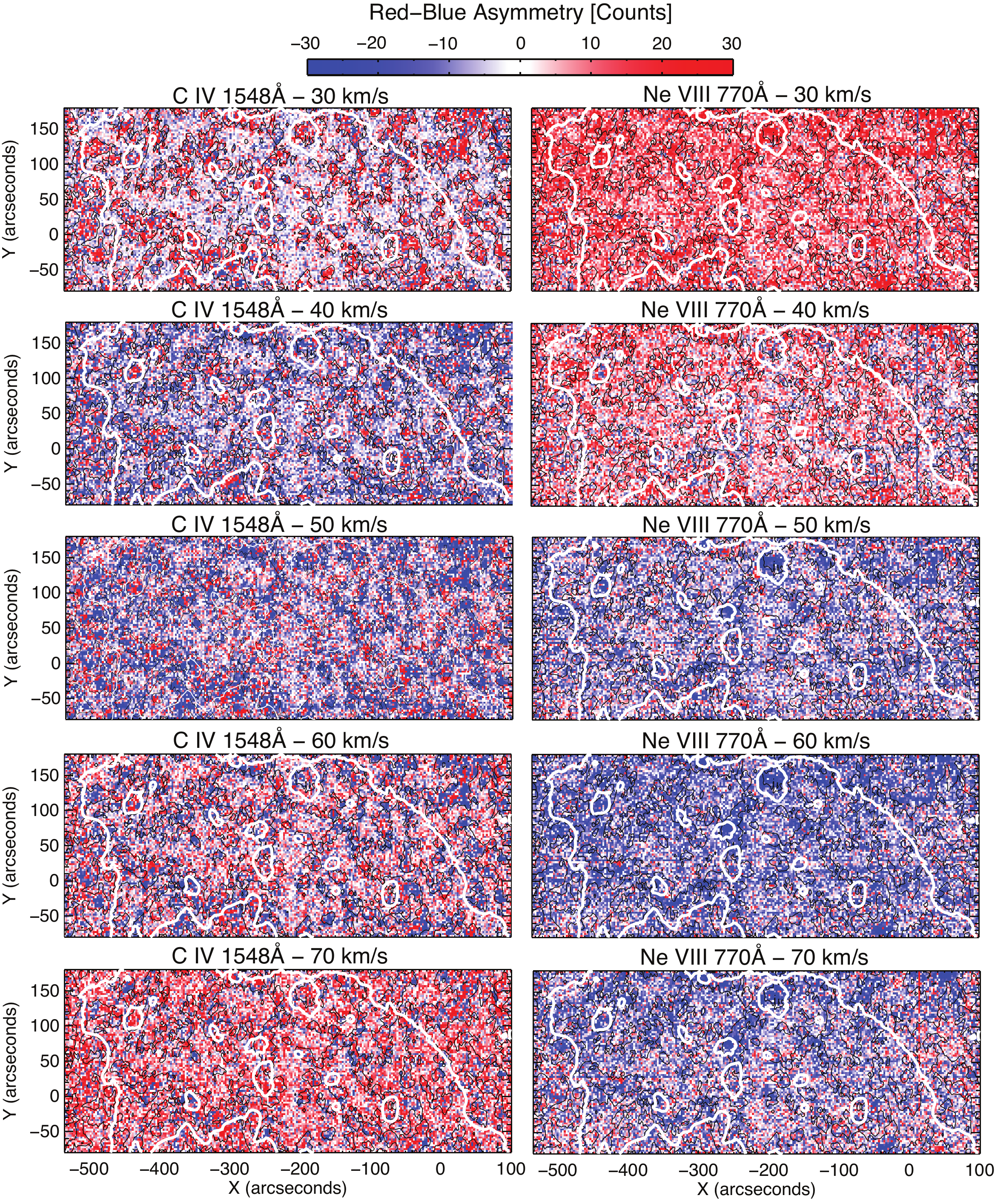}
\caption{Red-blue (R-B) profile asymmetries for the \ion{C}{4} 1548~\AA{} (left) and \ion{Ne}{8} 770.4~\AA{} (right) emission lines of the November 6 1999 SUMER observations. Again, for reference, the panels of the figure are overlaid with the 150 count intensity contour of the \ion{C}{4} intensity as a proxy for the supergranular network boundary (thin black line) and the 150 DN EIT 195\AA{} intensity contour as the ECH boundary (thick line). The electronic edition of the journal has movies showing the panels of this figure along with those at intermediate velocities and the components of Fig.~\pref{f2}. \label{f9}} 
\end{figure*}

\begin{figure*} 
\plotone{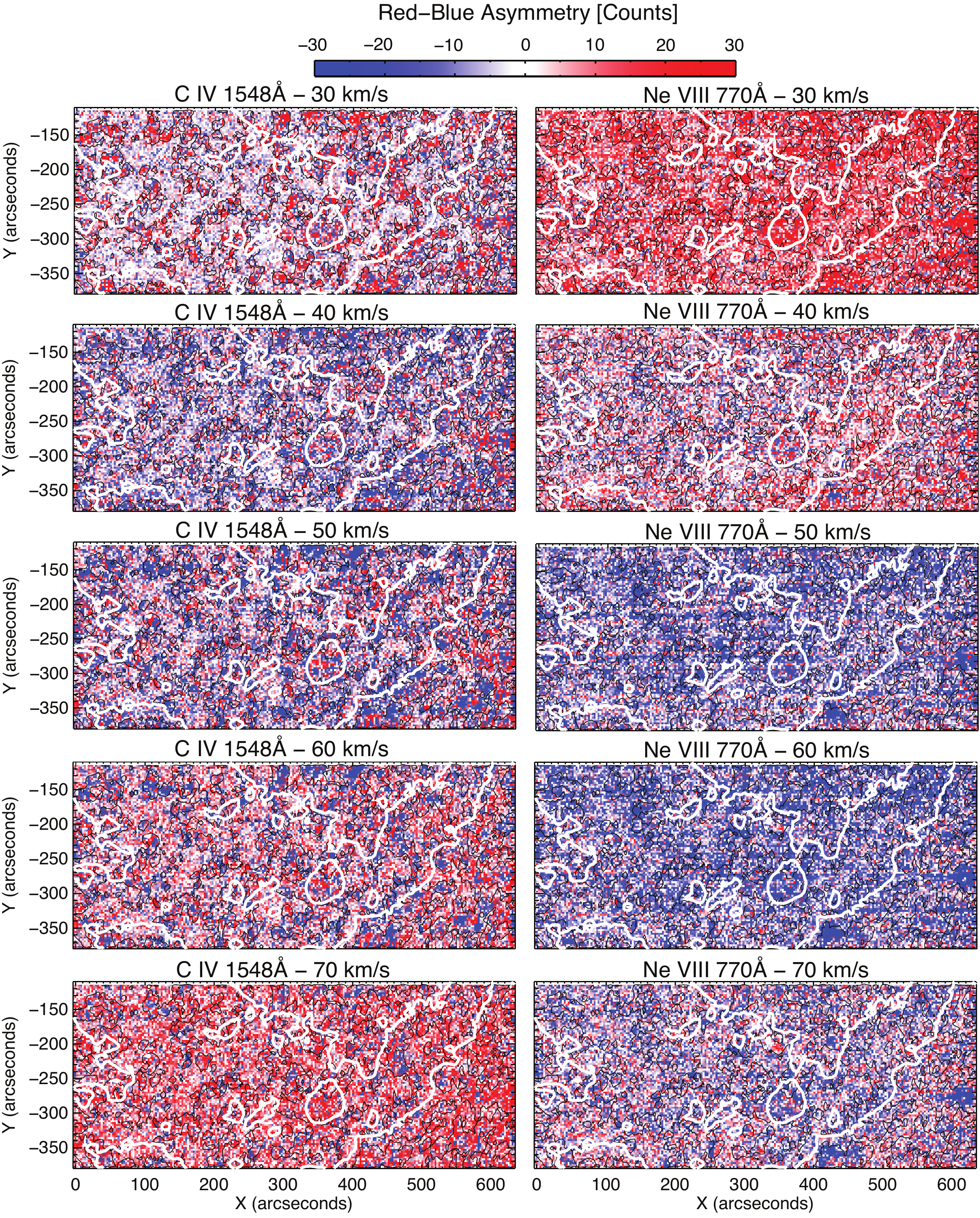}
\caption{Red-blue (R-B) profile asymmetries for the \ion{C}{4} 1548~\AA{} (left) and \ion{Ne}{8} 770.4~\AA{} (right) emission lines of the November 6 1999 SUMER observations. Again, for reference, the panels of the figure are overlaid with the 150 count intensity contour of the \ion{C}{4} intensity as a proxy for the supergranular network boundary (thin black line) and the 150 DN EIT 195\AA{} intensity contour as the ECH boundary (thick line). The electronic edition of the journal has movies showing the panels of this figure along with those at intermediate velocities and the components of Fig.~\pref{f3}.\label{f10}} 
\end{figure*} 

\subsection{Assessing the Symmetry of the Line Profiles}\label{sanal4}

Extending our study of the non-thermal broadening we need to assess the degree of profile symmetry, or asymmetry, observed in the ECH using the technique discussed in \citet{DePontieu2009} and \citet{McIntosh2009b}. These papers tied the appearance of high-velocity, blue wing asymmetries of hot UV and EUV emission lines to the mass loading of the active region, coronal holes, and the quiet corona that often seem to be associated with chromospheric ``Type-II'' spicules \citep[][]{DePontieu2007b} observed with the Solar Optical Telescope \citep[SOT;][]{Tsuneta2008} of \hinode{} \citep{Hinode}. These tall spicules inhabit predominantly unipolar flux concentrations, have an apparent velocity between 40 and 150km/s, a dynamic timescale of 50-100s \citep[much faster than the 20-40km/s of ``classical'' spicules that have 3-10 minutes timescales, e.g.,][and many others]{Roberts1945, Beckers1968}, and are likely caused by magnetic reconnection \citep[][]{DePontieu2007b}. Type-II spicules exhibit only upward motion, with the feature fading along its entire length in a matter of seconds, suggestive of rapid heating to TR temperatures, a feature exhibited in the analysis of \citep[][]{McIntosh2009c}. Such heating is also suggested in so-called rapid blueshifted events that are thought to be the disk counterparts of type-II spicules \citep{Rouppe2009}.

As in \citet{McIntosh2009b} we perform a single Gaussian fit (with a constant background) to the line profile in each spectroheliogram pixel to determine its centroid position. Then we interpolate the line profile to a wavelength scale that is ten times finer and calculate the total number of counts in the interpolated profile in two narrow range of wavelengths/velocities (two original SUMER spectral pixels wide, or 24 km/s at 1548~\AA) that are symmetrically positioned from the line centroid position, one in the blue and one in the red wing. The difference of the counts in the two wing positions, red-blue (R-B), provides a measure of the asymmetry of the profile at the chosen wavelength or velocity offset. Of course, a measure of zero indicates that the profile is symmetric in that velocity range. The R-B measure is related to the third (skewness) and fourth (kurtosis) moments of the the line profile and is a robust measure of asymmetry, provided that no spectral blends are present in the lines, an issue that we will return to below.
 
Applying the R-B asymmetry measure to the two emission lines we produce the series of images shown in Figs.~\pref{f9} and~\pref{f10} (and accompanying movies) for a range of reference velocities in the two portions of the ECH observation. In both figures, and in both spectral lines at low velocities (below sound speed expected for the relevant formation temperatures), we see that the R-B asymmetry is dominated by the red wing. This dominance is clear in the supergranular boundaries, increasing further over the strongest magnetic flux concentrations that comprise the supergranular vertices. Moving to higher velocities, we see that the R-B asymmetry in the supergranular network switches sign - indicating that the blue wing of the line profile is becoming dominant, which is compatible with a preponderance of high-velocity upflows in the strong magnetic field regions. The change in sign is rapid for \ion{C}{4}, and occurs about 10-20km/s higher for \ion{Ne}{8}. Comparing the R-B asymmetry for \ion{C}{4} at 50km/s and \ion{Ne}{8} at 70km/s we see a striking similarity between the two panels. Even though the supergranular network pattern is not obvious in the \ion{Ne}{8} line intensities, the R-B map at 70km/s shows that the blue wing asymmetries are strongly correlated with the locations of network vertices. The behavior of the R-B maps is identical to those derived from quiet sun regions in these lines, plus several others \citet{McIntosh2009b}. We also note that there is no clear delineation of the coronal hole / quiet sun boundary in the R-B asymmetry maps of Figs.~\pref{f9} and~\pref{f10}. The lack of a clear quiet sun / coronal hole boundary has significant implications for the mass and energy transport in the upper solar atmosphere as we will discuss below.

\begin{figure} 
\epsscale{1.15}
\plotone{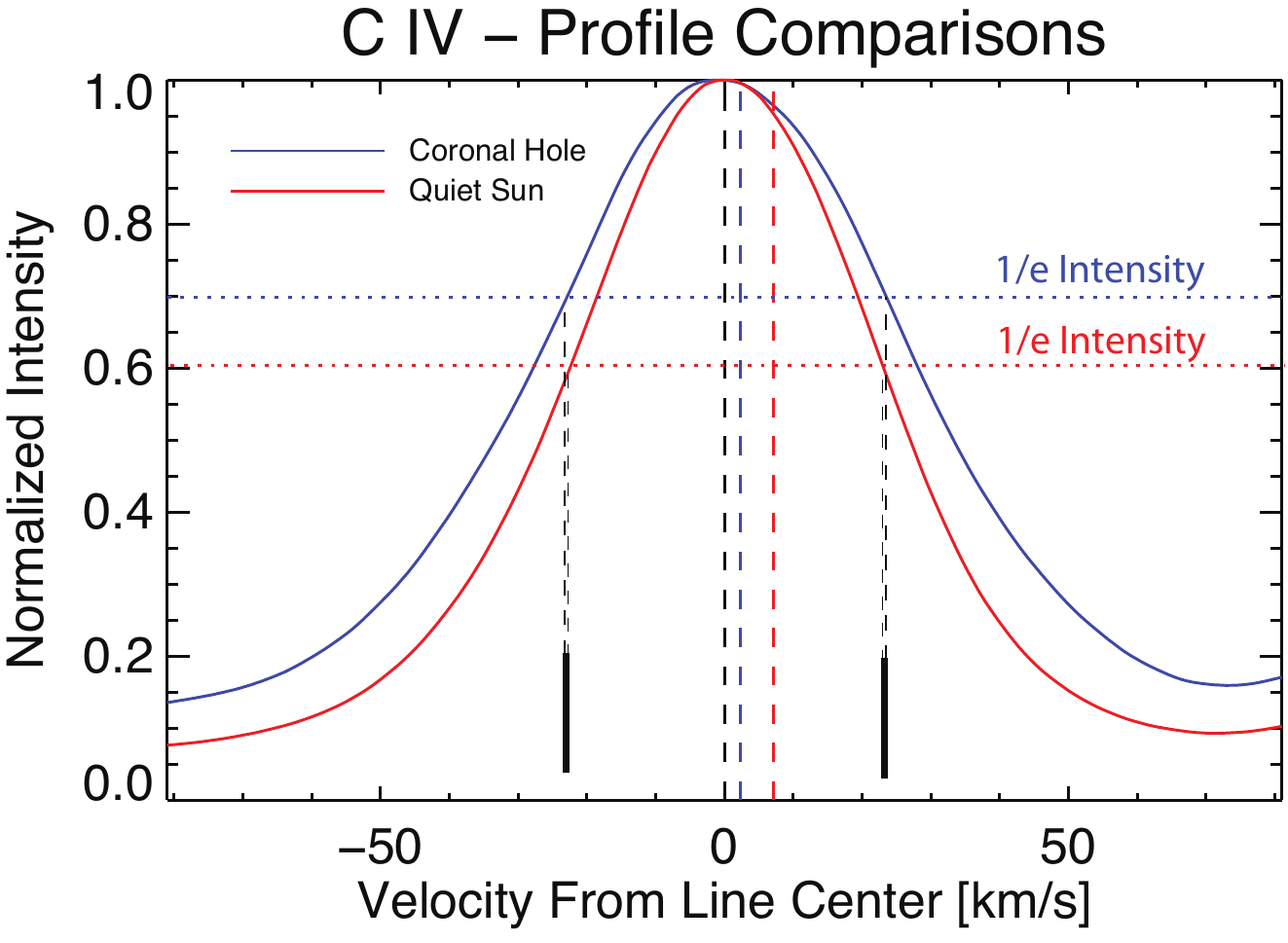}\\
\plotone{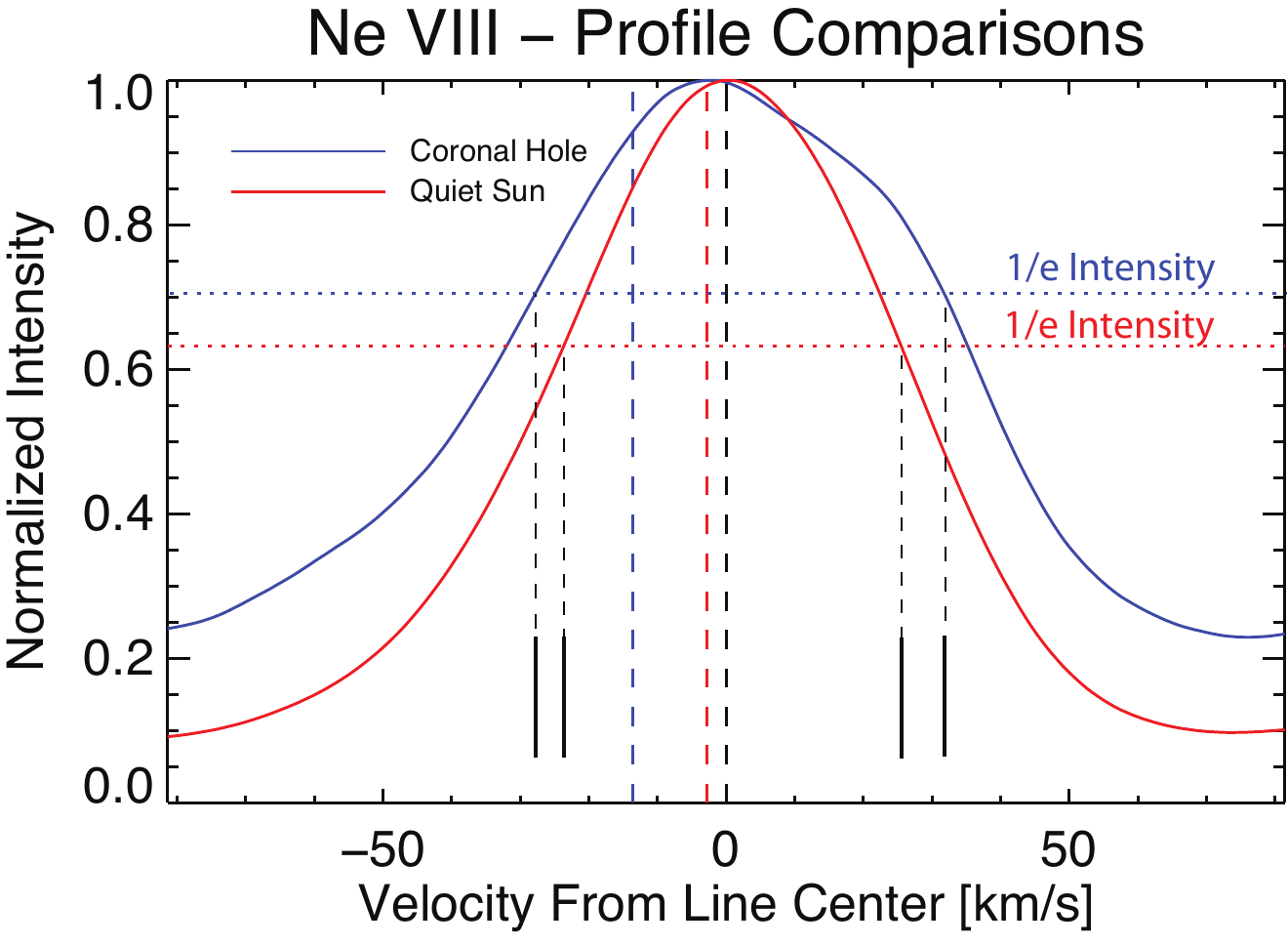}
\caption{Representative line profiles for \ion{C}{4} (top) and \ion{Ne}{8} (bottom) in a coronal hole (blue) and quiet sun (red) supergranular network vertex. In each panel we have placed the center-of-mass of the line profile at zero (black vertical dashed line) while the absolutely calibrated velocities of the profile in the coronal hole and quiet sun are shown as the blue and red dashed lines respectively. The correspondingly colored horizontal dotted lines mark the 1/e intensity level in each case from which we can determine the 1/e width of the line profiles with the vertically dropped dashed and solid black lines. \label{f11}} 
\end{figure} 
 
\subsection{Consolidating the Spectroscopic Information}\label{sec:cons}

In an effort to understand the relationships shown above, Fig.~\pref{f11} shows example (normalized; $I/I_{max}$) line profiles for magnetic network elements in the coronal hole (blue solid line; located at 420\arcsec, -325\arcsec) and quiet sun (red solid line; located at 200\arcsec, -250\arcsec) in \ion{C}{4} (left) and \ion{Ne}{8} (right). 

In the case of \ion{C}{4} we can immediately understand why the values of $v_{nt}$ are so similar between the quiet sun and coronal hole network; the relative reduction of the coronal hole network peak intensity (which is the total intensity minus the continuum intensity in Fig. \pref{f11}) is of the order of 5\% (small, assuming that the continuum emission is not significantly affected between quiet sun and coronal hole) and pushes the 1/e intensity level upward compared to the quiet sun profile. This compensates for the shallower gradient (with wavelength) in the wings of the coronal hole line which {\em should} increase the measured values of  $v_{nt}$. Combining these two effects results in only a small difference in the dropped black lines at the 1/e intensity level of each profile. We compare the blue and red vertical dashed lines which, again, show the relative shift to the blue of the  \ion{C}{4} coronal hole line profile. In the parlance of \citet{McIntosh2009b} this shift of the profile blueward may be caused by a (significant) reduction amount of the lower TR plasma that is cooling out of the corona, in the magnetically open network compared to that of the closed quiet sun network. This is compatible with recent results from 3d MHD simulations that suggest that the pervasive redshifts in the transition region are related to plasma draining from the corona after the field lines are disconnected from the energy injection site \citep{Zacharias2009}. However, conflicting results have been obtained in recent numerical 3d MHD simulations which suggest that the pervasive redshifts in the transition region may be a natural consequence of episodic heating events in which upper chromospheric plasma is heated to coronal temperatures at low heights. In the latter scenario, the subtle difference in \ion{C}{4} redshifts between coronal hole and quiet Sun may be caused by a difference in the distribution of heights and amplitude of the heating events \citep{Hansteen2010}. Further numerical simulations for varying magnetic field configurations will be required to help understand the observed differences.

Given the visual appearance of the \ion{Ne}{8} data (Figs.~\pref{f3}, \pref{f6}) and the scatterplots (Fig.~\pref{f7}) we would expect that the differences between profiles of the quiet sun and coronal hole network are considerable, and they are. Perhaps the most dramatic difference is the shoulder in the red wing of the line, but this doesn't significantly alter the results presented here. We point the interested reader to Appendix~A where we explore the curious shape of the ECH \ion{Ne}{8} line profile in a little more detail discussing the impact of \ion{Si}{1} blends on the line shape, derived R-B measure, and $v_{nt}$ diagnostics. The significant reduction in the \ion{Ne}{8} peak intensity in the coronal hole network (a factor of $\sim$2.5) could be associated with several effects. For example, it may be indicative of a significantly reduced contribution to the line profile from material returning from the overlying open corona, i.e., a large portion of the material that gets heated in the network to the formation temperature of \ion{Ne}{8} (or higher) is injected into the fast solar wind. On the other hand, the open field conditions are expected to lead to very different contributions to the energy balance from the thermal conductive flux and enthalpy flux, which would lead to a different temperature stratification and thus TR emission measure distribution with temperature \citep{Hansteen1995}.

A significant reduction in returning mass from the open corona would profoundly impact the line profile. This is compatible with what we observe. We see a large (10km/s) shift of the line centroid to the blue, as the balance of mass-transport shifts to predominantly upflowing material. Further, this reduction in intensity of the core profile component will compound with the net increase of the blue-wing asymmetry relative to it (assuming that it does not change considerably in magnitude in a coronal hole, as we have seen). This combination will naturally lead to a noticeably (5-10km/s) larger 1/e width of the single Gaussian fit to the profile. These points again illustrate the perils of using only the first (line center) and second (line width) moments of the profile to infer the properties of a truly dynamic transition region plasma, as noted in the conclusion of \citet{McIntosh2009b}. 

We also notice that the coronal hole profiles of \ion{C}{4} and \ion{Ne}{8} both show gradients in their wings that are systematically shallower than their quiet sun counterparts. This is indicative of an additional broadening source. We speculate that this symmetric excess component of the line-of-sight broadening is the result of the transverse (\alfvenic) motion of the sub-resolution motion of the type-II spicules that exist in these magnetic regions \citep[][]{DePontieu2007, McIntosh2008}. Based on the measurements of \citet{DePontieu2007b} the spicules are mostly vertical, but have a mean inclination of 20-30 degrees from the normal, and for a transverse wave amplitude of 10-20km/s \citep{DePontieu2007,Rouppe2009} this would produce a contribution in the line of sight of $\sim$3-7km/s which could increase to $\sim$5-10km/s in the rarer low-coronal plasma. Of course, in a coronal hole, the rapid density drop of the plasma with height would produce systematically larger amplitude wave motions \citep{McIntosh2009a} in both lines, furthering the symmetric broadening of the profiles by a few km/s. 
The \alfvenic{} motions may also help explain the enhancements of $v_{nt}$ that straddle the network boundaries in \ion{C}{4} (Figs.~\pref{f2},~\pref{f3}). It is tempting to speculate that the sub-resolution \alfvenic{} motion of the spicules on the larger inclination of the field spanning out into the internetwork will give rise to much larger values $v_{nt}$ than elsewhere \citet{McIntosh2008}.

\section{Discussion}\label{results}
 
We have studied the line broadening and profile asymmetry of the \ion{C}{4} 1548~\AA{} and \ion{Ne}{8} 770.4~\AA{} emission lines in a large equatorial coronal hole as an extension of the analysis presented in  \citet{McIntosh2006, McIntosh2007}. We see that there are subtle changes in the \ion{C}{4} network line profiles in an equatorial coronal hole over those in the quiet sun: the line-center red-shift of the emission is reduced by $\sim$4km/s; the peak emission is dimmer by 5\% and; the line wings have a shallower gradient. These effects contribute, as we have seen above, to an apparently invariant network $v_{nt}$ between coronal hole and quiet network regions. For the low corona\footnote{Technically, \ion{Ne}{8} could also be considered upper transition region emission. However, we believe \ion{Ne}{8} emission is more closely tied to coronal structure than that of the transition region, given the significant reduction in cell-interior intensity contrast for \ion{Ne}{8}} we see: strongly reduced intensities; blue-shifted line centers; and excessive line broadening in the coronal hole compared to the quiet sun. For the first time, we also analyze in detail the pervasive blue wing profile asymmetries in and around the network in a coronal hole. We have found these blueward asymmetries in a wide variety of spectral lines that span the entire transition region: \ion{C}{4} 1548\AA{} and \ion{Ne}{8} 770\AA{} (in second order), as well as \ion{O}{6} 1031\AA, and \ion{Ne}{8} 770\AA{} in first spectral order (see Appendix~B).  

Our observations of the dependence of intensity, line centroid, width and asymmetries as a function of temperature and magnetic field configuration (open vs. closed) provide strict constraints to theoretical models aimed at explaining the structure and properties of the transition region. Such constraints are crucial for these models: while they are rapidly advancing in complexity, with both 3d MHD models \citep{Zacharias2009,Hansteen2010} and Sun to Earth 1.5D models becoming available \citep[e.g.,][]{Suzuki2006,Cranmer2007}, significant discrepancies and inconsistencies continue to exist, with some models providing contradictory explanations for, e.g., pervasive redshifts in the transition region (see \S~3.3).

None of the current models provide a compelling scenario that can explain the blueward asymmetries in and around the network. Based on previous observational work, we have proposed a scenario in which heating events low in the atmosphere (chromosphere or transition region) above network regions accelerate and heat plasma from chromospheric or low TR temperatures up to coronal temperatures, for the quiet corona and fast solar wind alike. We speculate that this mass transfer may be driven from below through magnetic reconnection. Depending on the exact height at which the non-thermal energy is deposited, we expect to see different diagnostic effects: if the energy is dumped at chromospheric heights, the upflows would be visible as Type-II spicules in the magnetic network of the chromosphere \citep{DePontieu2009, McIntosh2009b}, whereas an event occurring at TR heights would lack a chromospheric counterpart, but nevertheless cause a faint, upflowing component in TR/coronal lines. Generally, this scenario of heating at low heights is consistent with a picture presented in \citet{Parker1991}. 

One very interesting observational finding is that we see no difference in the magnitude of the asymmetry of the line profiles upon crossing from the open to closed magnetic regions. The apparent invariance of the asymmetry in open and closed magnetic region has implications for the relative importance of magnetic heating at coronal ``heights'' \citep[e.g.,][]{Klimchuk2006, Patsourakos2006} compared to energy released in the lower atmosphere. Based on the arguments of \citet{Parker1988} and \citet{Parker1989}, were the chromospheric mass being driven upward as a result of energy release at coronal heights, the signature of the mass transport seen in the lower atmosphere of a magnetically open region should look quite different to that in closed field regions. This is because the open coronal magnetic field is expected to inhibit the occurrence of coronal field-line braiding. In addition, the changing mix of thermal conduction and enthalpy flux in the energy balance in open vs. closed field regions would be expected to also significantly affect the magnitude of the high velocity upflows in coronal holes \citep{Hansteen1995}. That change would lead to a pronounced ``boundary'' in the R-B analysis when crossing from closed to open field. However, this is {\em not} observed.
 
Like that presented in \citet{DePontieu2009} and \citet{McIntosh2009b}, the scenario we propose for the small-scale heating/upflow events is related to the model put forth by \citet{Pneuman1978} who suggested that a small fraction of the mass in chromospheric spicules is heated to coronal temperatures - providing the corona with hot plasma. 
Following the example of \citet{McIntosh2009b}, and assuming that the upward mass flux in the quiet sun and coronal hole regions are identical (1-4 type II spicules of density 10$^{10}$cm$^{-3}$ per arcsec$^2$ per second in a magnetic network element), the spicular mass flux into the solar wind is sufficient to balance the observed mass-loss even if only 0.5-2.5\% of the chromospheric spicule reaches temperatures greater than 600,000K \citep{Athay1982}.

As we have stated above, the impact of these mass-loading events requires much more detailed study using new observing tools such as the Atmospheric Imaging Array (AIA) on the {\em Solar Dynamics Observatory} and the upcoming Interface Region Imaging Spectrograph (IRIS). Such follow-on studies will be crucial to determine the potential impact of these events on the composition and speed of the fast solar wind. For example, we do not know whether the temperature of the plasma in these high velocity upflows is significantly in excess of the equilibrium formation temperatures \citep{DePontieu2009, McIntosh2009d}. 

Numerous challenges to our understanding arise as a result of the presented material. If the temperatures reached by the spicules is far in excess of 1~MK  \-- readily visible in {\em SDO}/AIA observations of ECH and QS \citep{DePontieu2010b} and early, high cadence, {\em STEREO} observations at the limb \citep{McIntosh2010b} \-- then we would expect that the significant thermal pressure ensures that the super-heated portion of the spicule material overcomes the Sun's gravitational pull to become a constituent of the solar wind rooted in the open magnetic network \citep[][]{Parker1958}. Are the \alfvenic{} motions associated with these spicular upflow events \citep{DePontieu2007} triggered by the quasi-periodic mass-loading at the bottom of the magnetic field lines, and do those waves represent the significant source of low-frequency wave energy needed to accelerate the plasma to a typical fast wind speed at heights beyond the first few Mm above the Sun's surface \citep[][]{Parker1991,Cranmer2005,Suzuki2005,Suzuki2006,Cranmer2007,Verdini2007,Verdini2010}? Finally, the most important question, what is the physical process that drives the Type-II spicules upward and heats a significant fraction of the mass that it contains to temperatures greater than one million kelvin?
 
\section{Conclusion}\label{conclusion}

The observed spatial and thermal dependence of intensity, line centroid, width and line asymmetries presented in this paper provide strict constraints to models for the roots of the fast solar wind. The transition region line profiles we observe are the result of an intricate balance of hot and cold material at a broad range of velocities. The newly discovered line asymmetries suggest that a significant portion of the energy release responsible for the transport of heated mass into the solar wind is provided by episodic heating and high speed upflow events that occur above the network at low heights (sometimes observable as type-II spicules). We suggest that detailed studies of the magneto-convective driving mechanism behind these heating events, and in particular the chromospheric spicules is essential. The magnitude, location, and timing of the energy release in the lower atmosphere will have considerable impact on the ionization balance and line formation. We anticipate that such an investigation will help us to understand the composition, initial acceleration, and nascent speed of the solar wind.
 
\acknowledgements
SWM thanks Marco Velli, Egil Leer and Tom Holzer for very illuminating discussions about this, and related, work. We are indebted to Klaus Wilhelm for comments on the manuscript and discussions regarding the work of the SUMER team with respect to the Si blend impact on the \ion{Ne}{8} line profiles presented. The material presented was supported by the National Aeronautics and Space Administration under grants to SWM and RJL issued from the Living with a Star Targeted Research \& Technology Program (NNX08AU30G, NNH08CC02C respectively). In addition, part of the work presented here is supported by grants NNX08AL22G, NNX08BA99G and NNX08AH45G to SWM and BDP. {\em SOHO} is a mission of international cooperation between ESA and NASA. Finally, we would like to acknowledge the (anonymous) referee who's comments strengthened the argument presented in this manuscript. The National Center for Atmospheric Research is sponsored by the National Science Foundation.

\clearpage
 
\appendix
\section{The Potential Impact of Spectroscopic Blends And Low Signal}\label{sec:blends}
\setcounter{section}{1}

\begin{figure} 
\epsscale{1.0}
\plotone{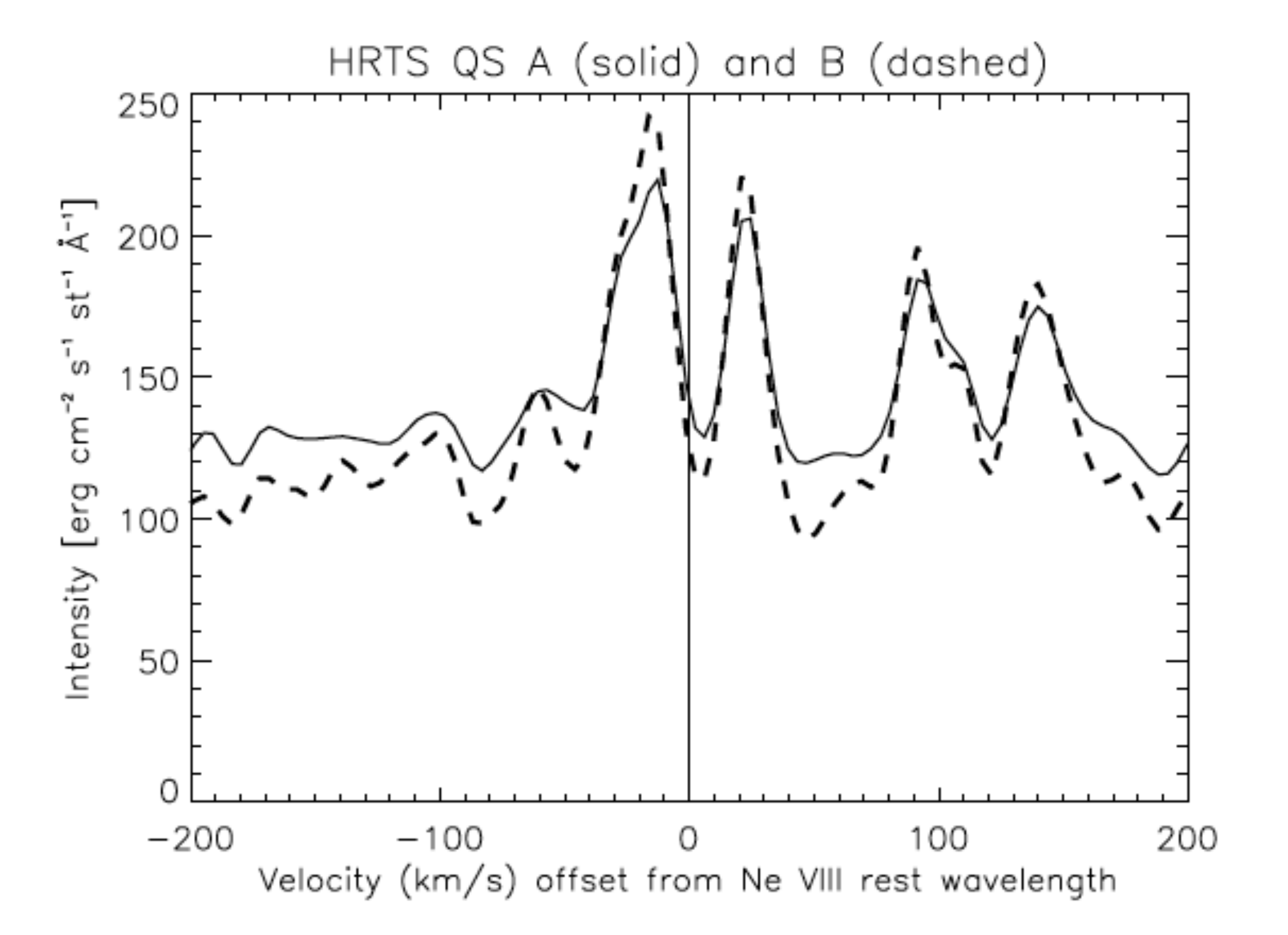}
\caption{Spectral atlas spectral profiles of two quiet Sun regions (full and dashed lines) from the HRTS rocket program shows two weak \ion{Si}{1} lines with wavelengths close to the \ion{Ne}{8} 770~\AA{} rest wavelength (observed with SUMER in second order at 1540.85~\AA{}). These blends are weak (2-5\% of peak quiet Sun \ion{Ne}{8} intensity), and at velocities of order 20 km/s from the \ion{Ne}{8} rest wavelength (vertical solid line) from \citet[][]{Davey2006}. \label{f12}} 
\end{figure}

In the right hand panel of Fig.~\pref{f11} we see an obvious shoulder in the red wing of the mean \ion{Ne}{8} 770.4~\AA{} profile in the ECH network. Is this the signature of a blend in the line profile and what impact does it have on the presented analysis? To estimate the impact of the blends on the $v_{nt}$ and R-B asymmetry maps derived in this Paper we use quiet Sun data from the HRTS spectral atlas \citep[][]{Brekke1993} since it only contains UV spectral lines observed in first order (i.e., it lacks the \ion{Ne}{8} line). We find that there are two \ion{Si}{1} blends (1540.74 and 1540.96\AA{}) that are of similar magnitude (of order 100~erg~cm$^{-2}$~s$^{-1}$~st$^{-1}$~~\AA$^{-1}$ above the continuum level) near the line center of the \ion{Ne}{8} line \citep[see Fig.~\pref{f12}, and Fig.~2 of][]{Hassler+others1999}. In quiet Sun, the \ion{Ne}{8} line as observed with SUMER in second order has an intensity of a magnitude between those of \ion{C}{4}1548~\AA{} and 1550~\AA{} \citep[][]{Davey2006}. To determine the relative size of the \ion{Si}{1} blends to that of the \ion{Ne}{8} in the SUMER data we compare the HRTS measurements of the \ion{Si}{1} blends with those of \ion{C}{4} 1548 and 1550\AA{}. The latter are at levels of order 2000 to 4000~erg~cm$^{-2}$~s$^{-1}$~st$^{-1}$~~\AA$^{-1}$ for quiet Sun (in HRTS). This means that in quiet Sun, the \ion{Si}{1} are at a level of of order 3\% of the quiet Sun intensity of the \ion{Ne}{8} as observed with SUMER in second order.

\begin{figure*} 
\epsscale{1.0}
\plottwo{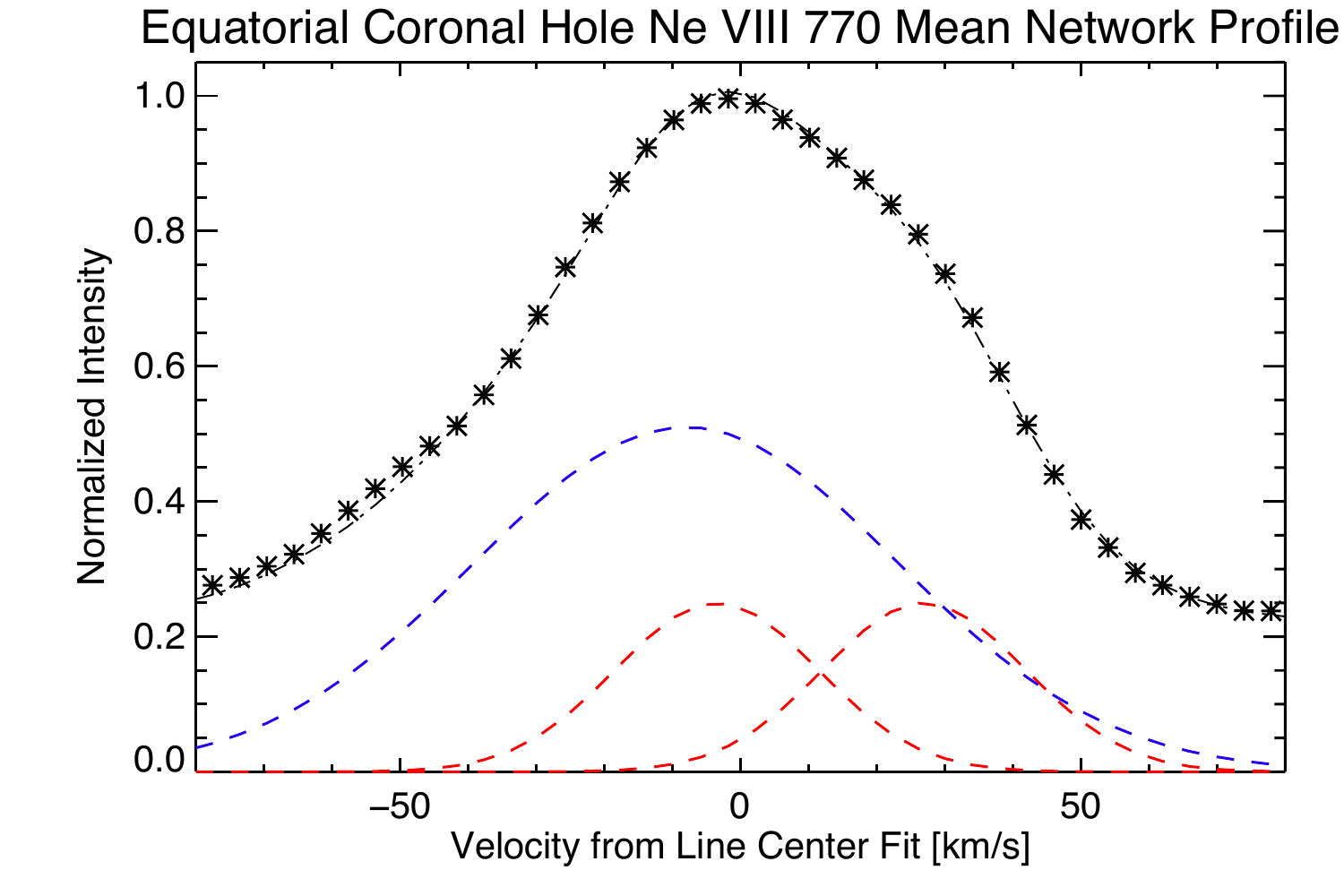}{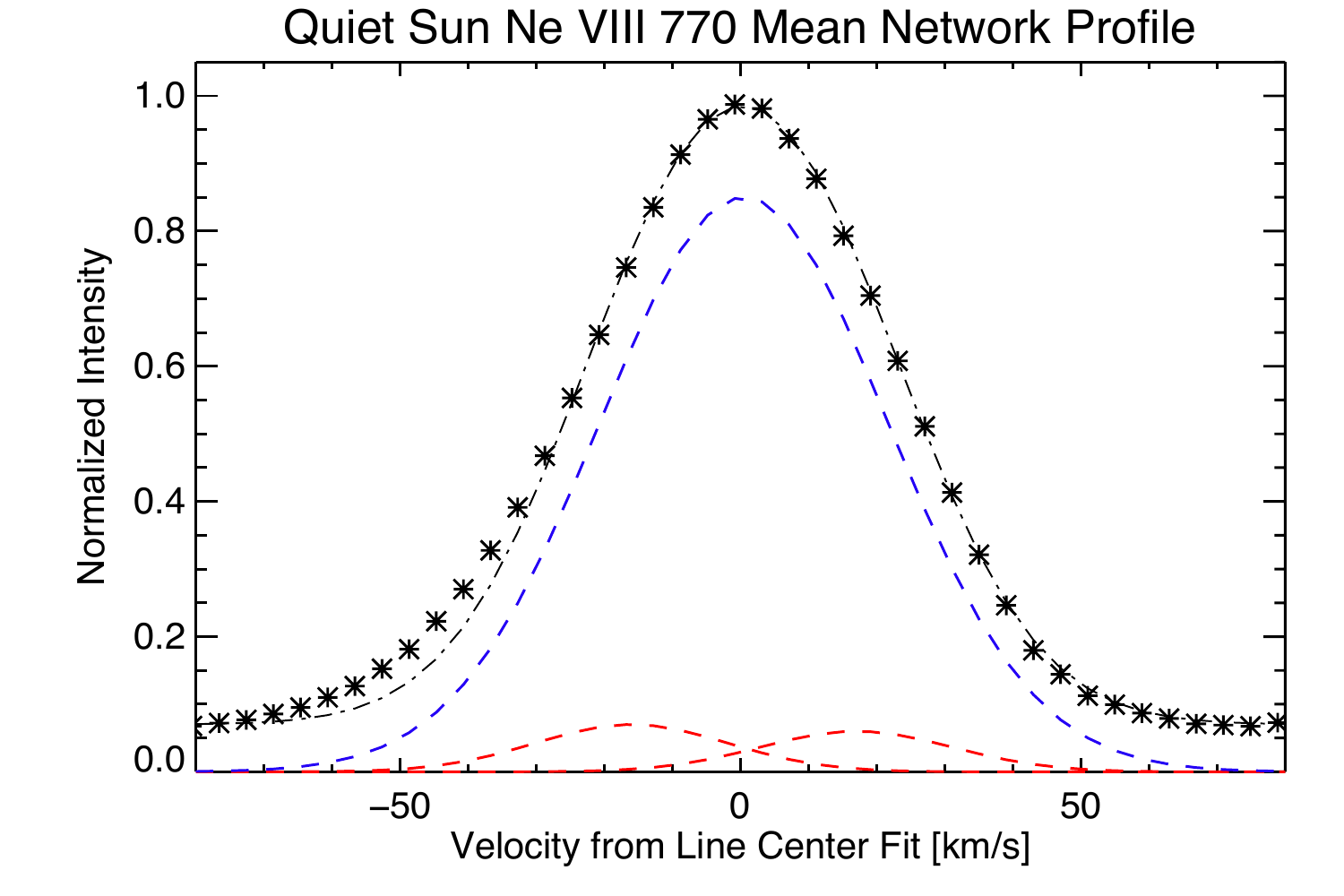}
\caption{Approximating the composition of the ECH (left) and quiet sun (right) network mean \ion{Ne}{8} 770~\AA{} line profile in second order using the two known \ion{Si}{1} blends from the HRTS reference spectrum in first order. For ease of visualization we have over-sampled the data (stars) by a factor of 5 in the spectral domain. The blue and red dashed lines represent Gaussian emission line profiles for \ion{Si}{1} and \ion{Ne}{8} respectively. Summed (with a constant continuum brightness) these comprise the total profiles (dot-dashed line). \label{f13}} 
\end{figure*}

In coronal hole network regions the \ion{Ne}{8} line has about a third of the quiet Sun network brightness, while the photospheric \ion{Si}{1} lines will have the same brightness irrespective of location. This means that, in the ECH, the \ion{Si}{1} blends should have a higher relative magnitude. We use these estimates to investigate the impact of these blends on our analysis of the data. In the two panels of Fig.~\pref{f13} we use a forward model that is based on a three Gaussian composite profile (two \ion{Si}{1} Gaussian lines of the same brightness separated by 36km/s and estimating the \ion{Ne}{8} Gaussian from the residual) and compare it with our average observed profiles in the ECH network (left) and quiet Sun (right). In both cases we find that the \ion{Si}{1} blends have the same intensity (above the continuum) as the continuum background itself.

For the ECH we see that the \ion{Si}{1} blends could be of order 10-20\% of the \ion{Ne}{8} line brightness, while in the quiet Sun they are at most a few percent. For the ECH, we see that the short wavelength \ion{Si}{1} blend and the \ion{Ne}{8} lie almost on top of one another with the longer wavelength blend creating the shoulder in the red wing. As a result, the centroid of the composite profile is shifted by 8~km/s to the red, compared with the centroid of the actual \ion{Ne}{8} line (21km/s). For the quiet Sun case, the two \ion{Si}{1} blends are almost symmetrically distributed about the \ion{Ne}{8} profile, which is blue-shifted by a couple of km/s. 

\begin{figure*} 
\epsscale{1.0}
\plottwo{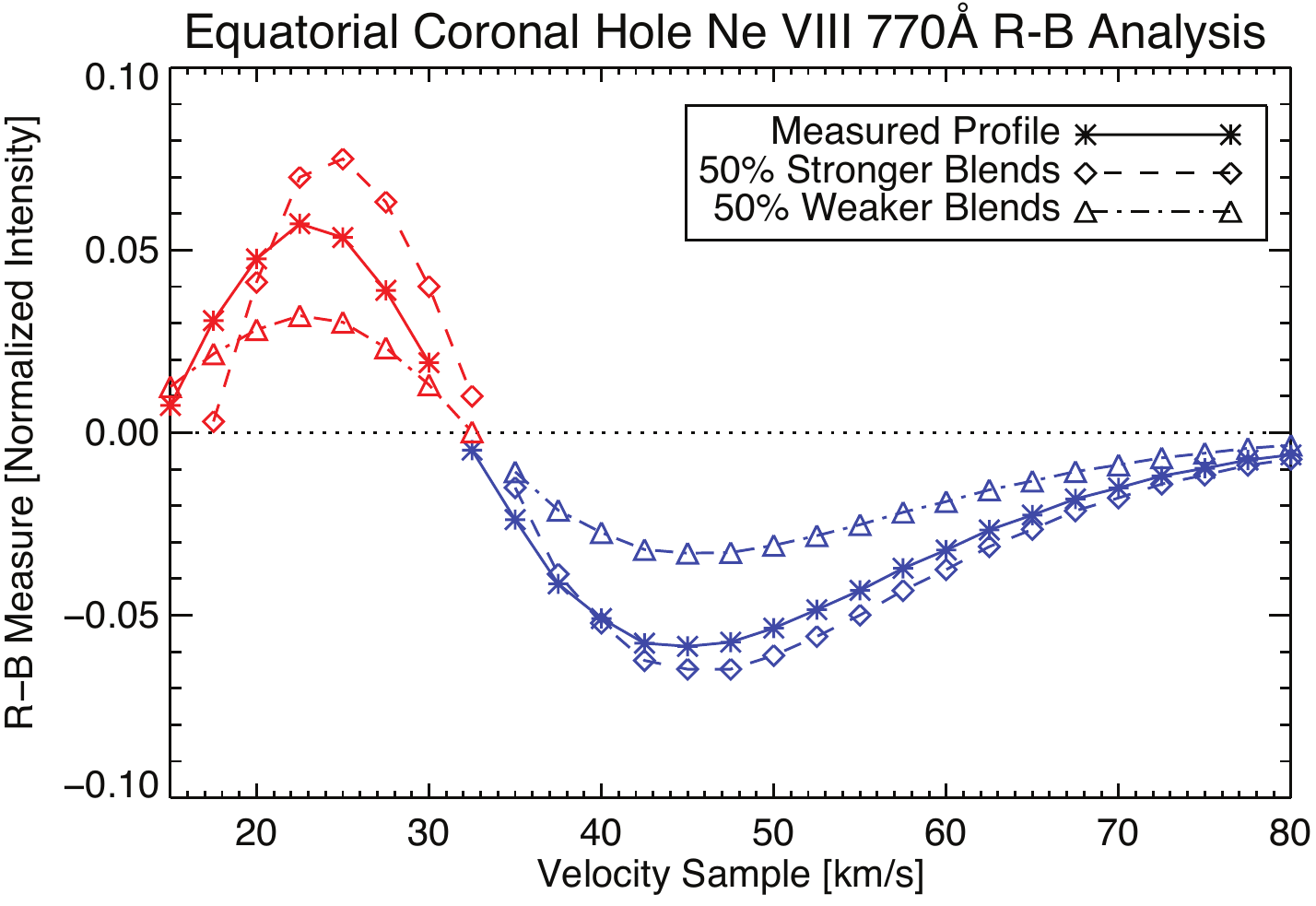}{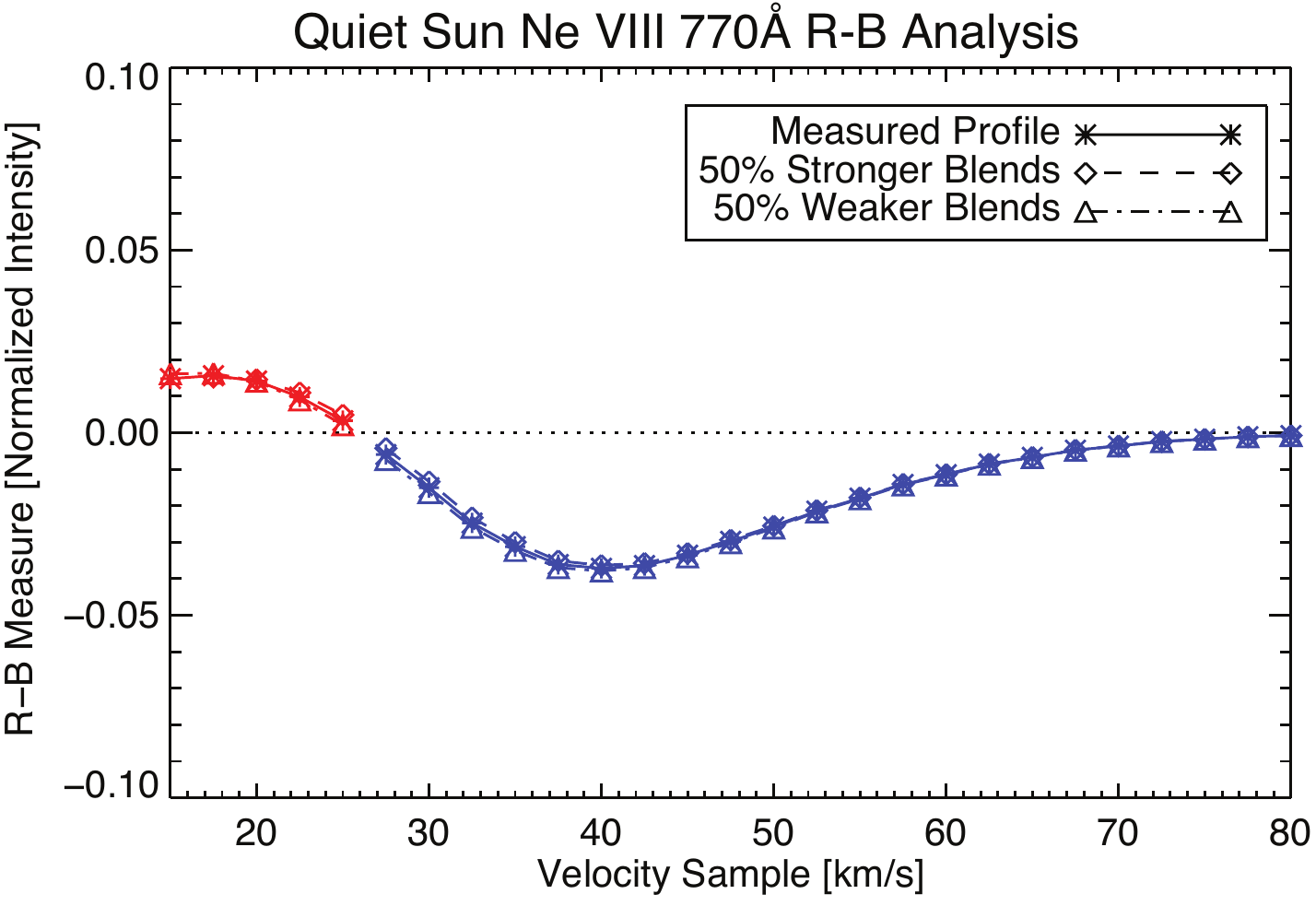}
\caption{Testing the R-B analysis of the ECH (left) and quiet sun (right) network mean \ion{Ne}{8} 770~\AA{} line profile using the position and magnitude of the \ion{Si}{1} blends as determined from Fig.~\pref{f13}. In each case we show the R-B profile as a function of the sampled velocity for the measured network profile ($\ast$), a case where the \ion{Si}{1} blends have 50\% larger ($\diamond$) and 50\% smaller ($\triangle$) amplitudes, respectively. The color used delineates if the R-B measure is positive (the signature of red wing asymmetry) or negative (asymmetric in the blue wing).  \label{f13b}} 
\end{figure*}

To illustrate the potential impact of these \ion{Si}{1} blends on the R-B measure used for \ion{Ne}{8} in the main body of the Paper we use the fitted component parameters for Fig~\pref{f13}. We have added the 5\% amplitude component centered on 60km/s in the blue wing, with a 1/e width of 25 km/s, that is needed to describe the measured profile, to conduct a simple experiment. In this experiment, for which the results for the ECH and quiet Sun profiles are shown in the left and right panels of Fig.~\pref{f13b}, we compare the R-B profile (as a function of velocity) of the measured network profiles ($\ast$) with those for which the \ion{Si}{1} blends are 50\% larger ($\diamond$) and 50\% smaller ($\triangle$) in magnitude than those determined from the corresponding panels of Fig~\pref{f13}. We see that the R-B profiles for the measured ECH and quiet Sun network profiles are very similar. The largest impact on the R-B analysis with modified blend amplitudes is on the ECH line profile, as the three quiet Sun profiles are indistinguishable. In the case of the ECH R-B profile we see that the main change is the magnitude of the asymmetry observed (of the order of a few percent) with the larger amplitude blends having a larger effect in the red wing (enhancing the profile shoulder), but little on the blue wing. This is somewhat different for the lower amplitude blends where there is a noticeable difference (2\%) in both wings. The velocity position of the blue asymmetry remains relatively stable (as we might expect since we have not moved the modified blends in velocity-space) shifting by a few km/s further to the red as the blends get weaker. This is also expected and is due to the decreased weighting of the red wing shoulder such that we are able to ``see'' the blue wing asymmetry at lower velocities. This experiment lends support to our belief that the \ion{Si}{1} blend configuration does not significantly impact the faint (2-5\%) asymmetry in the blue wing of the profile in either case, although the blends and profile shape affect centroiding the raw profile accurately. In quiet Sun, the strength of the \ion{Ne}{8} line compared to that of the blends makes the R-B analysis more robust. The weak, and roughly symmetrically positioned, blends around the \ion{Ne}{8} rest wavelength will mostly cancel each other in R-B asymmetry maps of the quiet Sun. In ECH, the incorrect centroiding of the \ion{Ne}{8} line profiles in the ECH by a few km/s to the red will skew the R-B asymmetry measure by a few km/s to the blue. For \ion{Ne}{8} this will result in the R-B maps having a blue hue, see Figs.~\pref{f9}, ~\pref{f10} and~\pref{f13b} where the R-B profile of the example does not return to zero. 

\begin{figure*} 
\epsscale{1.0}
\plotone{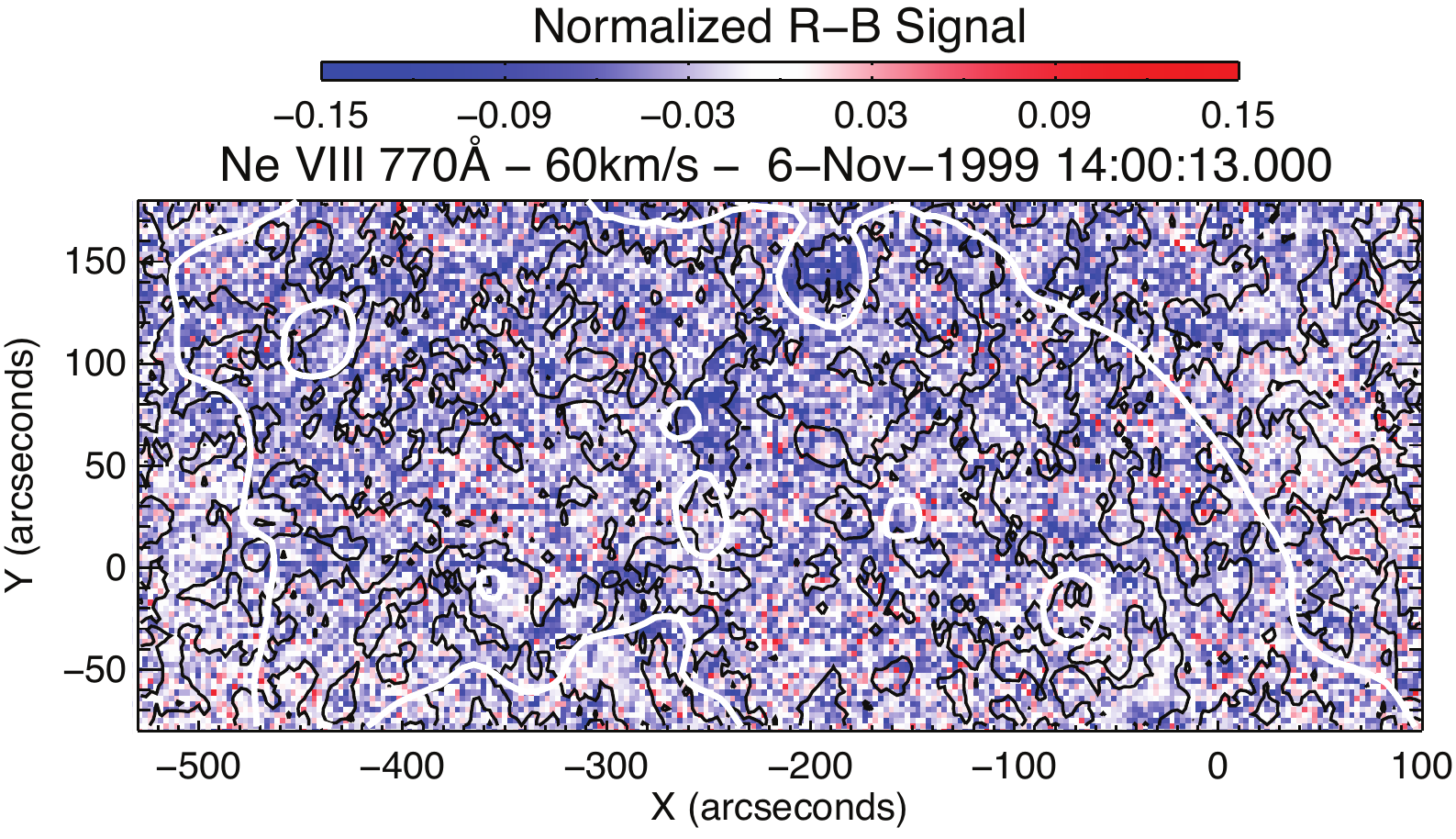}
\caption{Peak Intensity normalized R-B asymmetry map for \ion{Ne}{8} at 60 km/s from the November 6 1999 spectroheliogram (see Fig.,~\pref{f9}), showing the ECH boundary (white solid line) as determined from EIT 195\AA{} and the network emission in \ion{C}{4} (black closed contours) at 150 count levels. \label{f13_normalized}} 
\end{figure*}

However, we now show that this blue ``hue'' (caused by the \ion{Si}{1} blends) cannot explain all of the blueward asymmetries we see in \ion{Ne}{8} in the ECH, especially the stronger blue asymmetries in the ECH network. If the blueward asymmetries were caused only by the effect of the blends on the centroid of the profile, then we would expect the normalized R-B measure to be of the same magnitude in the network and in the internetwork. From Fig.~\ref{f13b} we know that stronger blends could, in principle, create a spurious excess blueward asymmetry. However, the network/internetwork contrast for the \ion{Ne}{8} is of order 2 \citep{McIntosh2007}, which is very similar to what can be expected for a chromospheric line such as \ion{Si}{1}. We note that the contrast for lines that are formed at similar temperatures like \ion{O}{1} and \ion{S}{1} is of order 1.5-2 \citep{Feldman1976}. As a result, the relative impact of the blends on \ion{Ne}{8} should be of the same size for network and internetwork, which would lead to ``flat'' normalized R-B map that shows no difference between network and internetwork. This is {\em not} what we observe: Fig.~\ref{f13_normalized} shows significantly larger blueward normalized asymmetries of \ion{Ne}{8} in and around the network, with many internetwork regions lacking asymmetric profiles.

\begin{figure} 
\epsscale{1.0}
\plotone{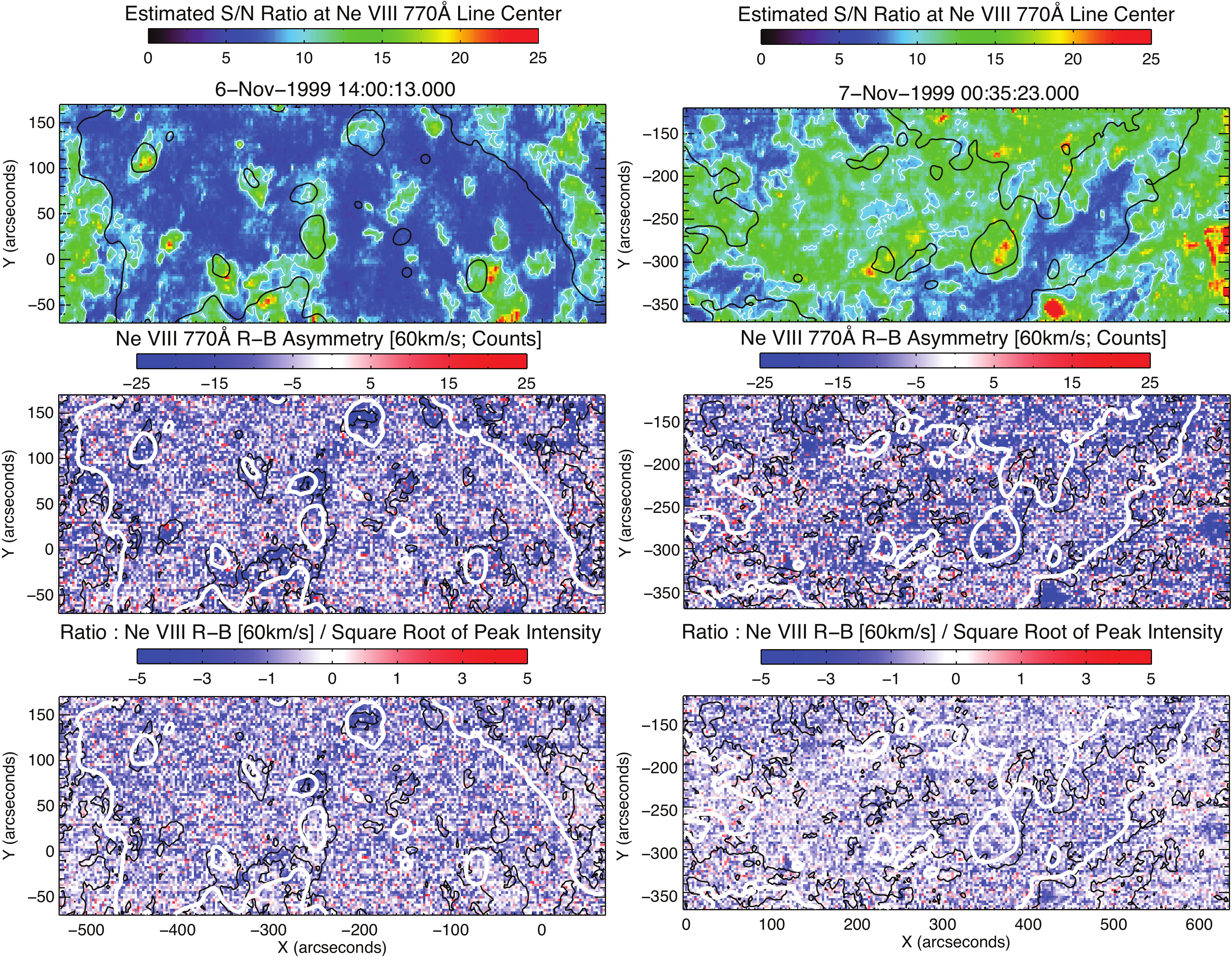}
\caption{Comparing the estimated signal-to-noise of the 9-pixel (3\arcsec $\times$ 3\arcsec) conglomerate \ion{Ne}{8} SUMER spectra (top) to the R-B profile asymmetry at 60km/s (cf. Figs~\pref{f9} and~\pref{f10}; middle), and their ratio (bottom), in the ECH observations of November 6 1999 (left column) and November 7 1999 (right column). An approximate \ion{Ne}{8} S/N level of 10 is shown in each panel as a thin contour (white \-- top row; black \--bottom). Similarly, the SOHO/EIT 195\AA{} coronal hole boundary is  shown as a thick black (top row), or white (bottom row), contour. \label{fs2n}} 
\end{figure}

Further, the individual SUMER spectra have relatively low count rates - the key motivating point behind our decision to add the spectra over the 3\arcsec $\times$ 3\arcsec{} region \citep[][]{Davey2006}. Low signal-to-noise in the line spectra adversely impacts the reliability of any estimate of the line profile asymmetry. To investigate this property of the \ion{Ne}{8} line spectra we construct an approximate measure of the signal-to-noise at each pixel in the form ${I_{peak}} / \sqrt{I_{peak}}$. Comparing this signal-to-noise estimate for each portion of the ECH observation to the R-B asymmetry maps (Figs~\pref{f9} and~\pref{f10}) in Fig.~\pref{fs2n} we see that regions of strong blue asymmetry generally correspond to regions where the signal-to-noise measure is larger than 10 (enclosed by contours). Those are places where we might logically expect the magnitude of the R-B asymmetry to be of order 10\% to be reliably detected. From the bottom panels of the figure we see that those regions of large profile R-B asymmetry are, at least, twice as large in magnitude as the square root of the peak line intensity at that pixel and so we assume that the determination of the profile asymmetry is real signal, and not the result of noise in the spectra. Based on the visual correspondence between regions of strong blue asymmetry and higher signal-to-noise shown in the figure we deduce that the blue-asymmetries observed are significant, albeit marginally above the noise threshold in the coronal hole, and are the signature of high velocity plasma upflows at those locations, as we may be led to believe by recent {\em SDO}/AIA observations \citep[][]{DePontieu2010b}, and the {\em TRACE} observations of this ECH that follow in Appendix~\pref{sec:trace_imaging}.

A final, independent confirmation in support of the ubiquity of the faint blue wing asymmetry in magnetic regions comes from more recent, lower signal-to-noise SUMER observations of quiet Sun which show the same faint blue-shifted component in \ion{Ne}{8} 770~\AA{} spectra when observed in first order \citep[i.e., without the \ion{Si}{1} blends][]{McIntosh2009b}. In \S~\pref{sec:fuckyou} we show that similar first order measurements of \ion{Ne}{8} 770~\AA{} {\em in ECH} confirm the existence of faint blue wing asymmetries. These first order measurements are immune to the effects of the \ion{Si}{1} blends. Further, the non-thermal broadening of the first order \ion{Ne}{8} (Fig.~\pref{f17}) is of a magnitude commensurate with those shown in earlier figures of the second order observations (Figs.~\pref{f2} -~\pref{f6}) adding further support to the fact that the \ion{Si}{1} blends do not sufficiently alter the diagnostic value of the measurements shown, consistent with the deduction of \citet{Wilhelm2000}.

We also consider potential blends in the \ion{C}{4} 1548~\AA{} line: we can clearly see in the 70km/s R-B asymmetry maps of Figs.~\pref{f9} and~\pref{f10} that there is a signature of another spectral line on the red side of the line. Based on the HRTS and SUMER \citep[][]{Curdt2001} spectral atlases, this is another \ion{Si}{1} blend. However, this blend is sufficiently far from the line center and weak that it does not impact the diagnostics presented other than to mask the true extent of the blue wing asymmetry. An unfortunate, but significant, casualty of these \ion{Si}{1} blends is the \ion{Si}{2} 1534~\AA{} line. It would have allowed for study of much cooler plasma, but the \ion{Si}{1} blend in the red wing of that line is place at about the 1/e width of the line from line center and would have significantly impacted both R-B and $v_{nt}$ diagnostics.

It is not clear what impact these \ion{Si}{1} blends have on other high-profile SUMER observations of the \ion{Ne}{8} line in second order and results derived from them \citep[see, e.g.,][]{Hassler+others1999,Dammasch+others1999,Tu+others2005a,Tu+others2005b}. The careful analysis presented in \citet{Wilhelm2000} concludes that the coronal hole \ion{Ne}{8} plasma, and the network line profiles in particular, in second order are {\em not} at the mercy of \ion{Si}{1} blends and that the network profiles are not singularly impacted by the Si blends. 

We note that caution is needed when interpreting these line profiles, there are at least two components contributing to the line profiles we observe and a single Gaussian fit to those profiles is fraught with danger \citep[e.g.,][]{McIntosh2009b}. We believe the only way to properly assess the impact of the \ion{Si}{1} blends is to perform a first/second order observation and compare the line spectra taken of the same region at approximately the same time \citep[e.g.,][]{McIntosh1998}. Unfortunately, such a test is not possible with SUMER and so we must rely on the first order diagnostics to illustrate the presence of the same spectroscopic characteristics as observed in the body of the present work.

\section{TRACE Imaging of the November 1999 ECH}\label{sec:trace_imaging}

\begin{figure} 
\epsscale{1.0}
\plotone{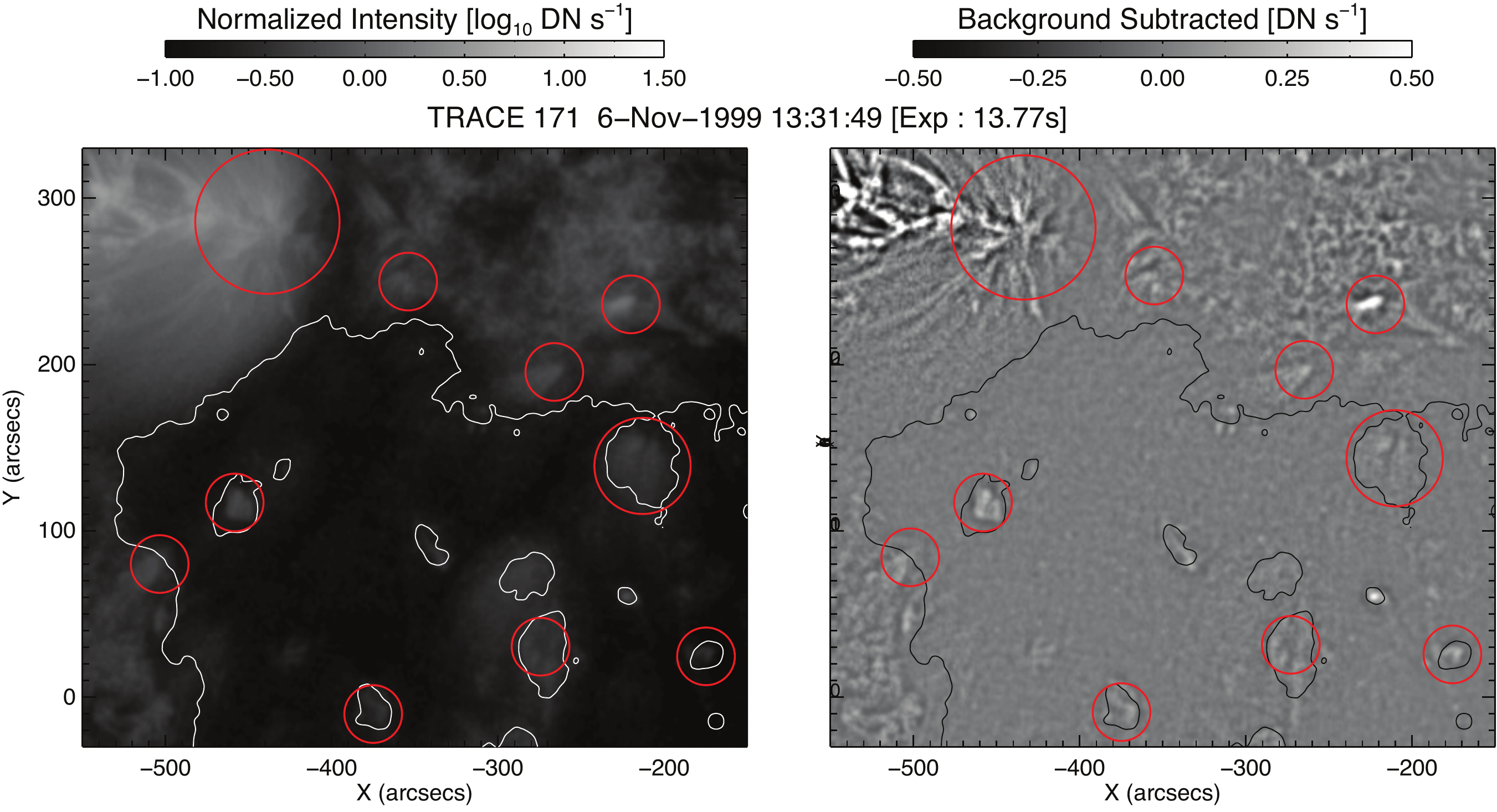}
\caption{The evolution of the ECH as observed by {\em TRACE} in the 171\AA{} passband on November 6 1999. The left panel shows a normalized intensity image while that on the right shows the same image with a smoothed (over 10 pixels \-- 5\arcsec) version of itself removed. The latter highlights fine-scale evolving structure. The online edition of the journal has a movie showing the temporal evolution of this region. The red circles highlight regions were propagating disturbances are visible, if weak, and dependent on the exposure time of the image. \label{trace}} 
\end{figure}

While SUMER observed the ECH the Transition Region and Coronal Explorer \citep[{\em TRACE};][]{Handy1999} observed it's north-west portion in the 171\AA{} passband. This passband consists of emission lines of \ion{Fe}{9} and \ion{Fe}{10}, i.e., it is sensitive to emission of plasma over 1MK. The exposure times for this sequence were variable due to the automatic exposure control of the telescope, but modulated between 3 and 14 seconds for the 12 hours of observation. The image shown in Fig.~\pref{trace} has an exposure time of 13.77s and the position of the coronal hole boundary in the field of view is shown as a contour familiar from many of the figures above. Inspection of the online edition movie supporting this figure suggests there are several regions where very weak propagating disturbances at coronal temperatures ($\ge$ 1MK) are visible \citep[similar to those described in][]{McIntosh2009d}.

These ``blobs'', or propagating disturbances, are readily and persistently seen in the active regions (top left of the field of view) and have similar speeds in other imaging experiments \citep[e.g.,][]{McIntosh2010a, McIntosh2010b, DePontieu2010b}. There are many other locations showing the same behavior (encircled in red) in the field of view, even some in the brighter emission regions of the coronal hole. Indeed, the majority (most clearly that centered on -210\arcsec, 140\arcsec) of these encircled locations are locations of strong blue asymmetries in the \ion{C}{4} and \ion{Ne}{8} lines (see, e.g., Fig.~\pref{f9}). Unfortunately, the erratic behavior of the {\em TRACE} exposure time doesn't permit the type of space-time plot analysis that we have performed for other {\em TRACE} \citep[][]{McIntosh2009d}, or {\em STEREO} observations \citep[][]{McIntosh2010b}, and so any further comparison between the {\em TRACE} image sequence and SUMER R-B analysis is purely qualitative. The sample figure and movie is included in this appendix for the sake of completeness and to illustrate the presence of propagating disturbances in some locations of \ion{Ne}{8} R-B asymmetries when the {\em TRACE} observation characteristics permit their observation.

\section{Additional Supporting SUMER Observations}\label{sec:fuckyou}
In this appendix we will present observations of line profile asymmetries in ECH from the \ion{O}{6} 1031.9~\AA{} \citep[cf. Sect.~2.3 of][]{McIntosh2009b} and in the unblended, first spectral order, \ion{Ne}{8} 770~\AA{} line profiles observed with SUMER's detector B. As noted in \citet{McIntosh2009b}, high ($\ge$20) signal-to-noise is essential to robustly measure these small amplitude profile asymmetries. Observations with the coated region of SUMER detector B have considerably higher signal-to-noise than the long-exposure detector A observations shown in the body of this Paper without the potential for impact on the spectra by cool, very weak blends.

\begin{figure} 
\epsscale{0.75}
\plotone{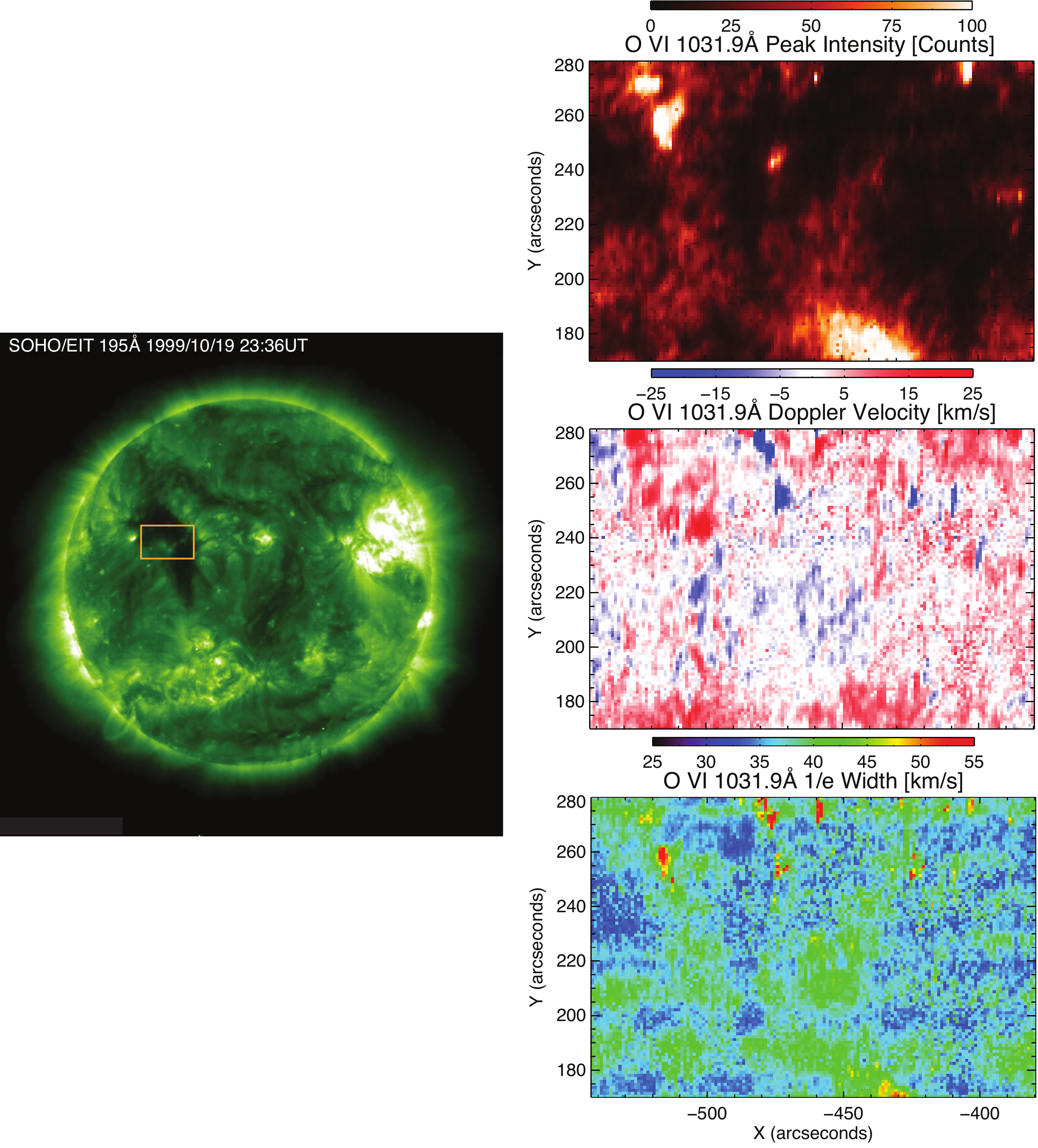}
\caption{The spatial variation of the SUMER spectroheliograms in the \ion{O}{6} 1031.9~\AA{} emission line in an equatorial coronal hole with 100s exposures on October 19-20 1999 (21:13-01:15UT). The left panel shows the EIT 195\AA{} image at the midpoint of the SUMER observation for context and the rastered region is indicated by the orange box. The panels on the right show the peak intensity of the line, the Doppler velocity, and the 1/e line width. \label{f13_o6}} 
\end{figure}

\begin{figure} 
\epsscale{0.75}
\plotone{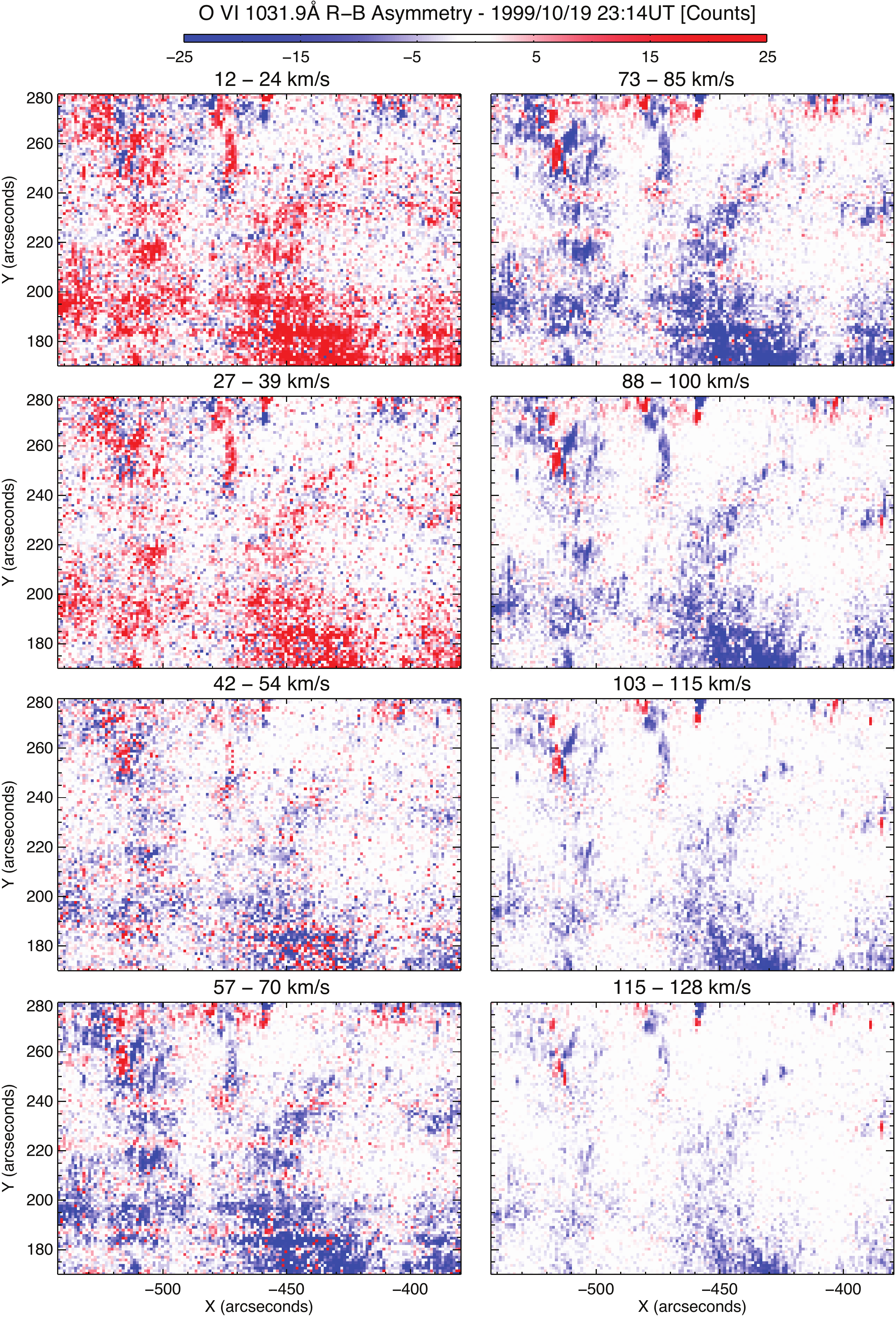}
\caption{Red-blue (R-B) profile asymmetries for the \ion{O}{6} 1031.9~\AA{} emission line at a range of velocities for the October 19-20 1999 (21:13-01:15UT) observation of an ECH. The electronic edition of the journal has movies showing the panels of this figure along with those at intermediate velocities and the components of Fig.~\pref{f13_o6}.\label{f14}} 
\end{figure}

\begin{figure} 
\epsscale{0.75}
\plotone{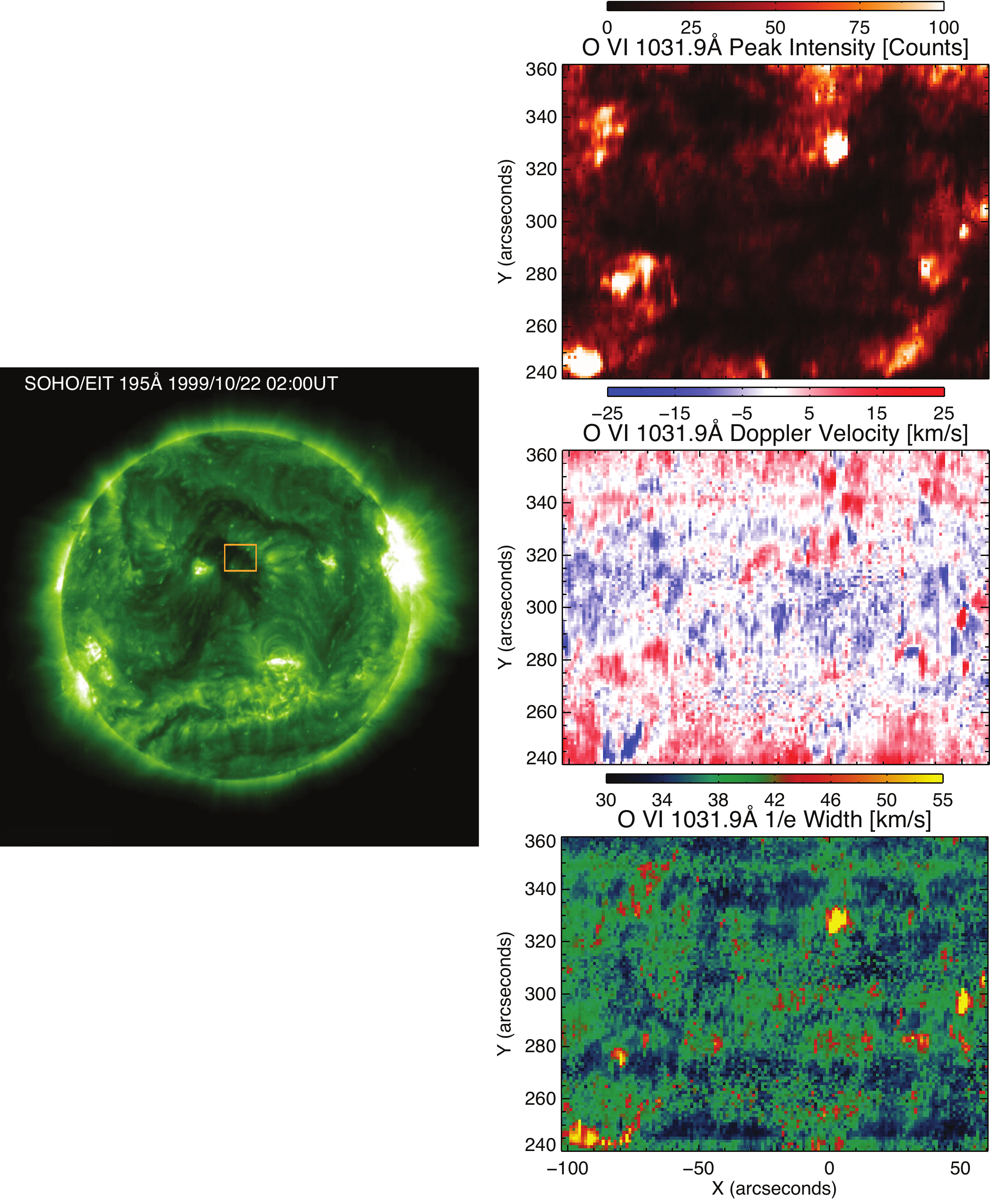}
\caption{The spatial variation of the SUMER spectroheliograms in the \ion{O}{6} 1031.9~\AA{} emission line in an equatorial coronal hole with 100s exposures on October 22 1999 (01:57-06:00UT). The left panel shows the EIT 195\AA{} image at the midpoint of the SUMER observation for context and the rastered region is indicated by the orange box. The panels on the right show the peak intensity of the line, the Doppler velocity, and the 1/e line width. \label{f15}} 
\end{figure}

\begin{figure} 
\epsscale{0.75}
\plotone{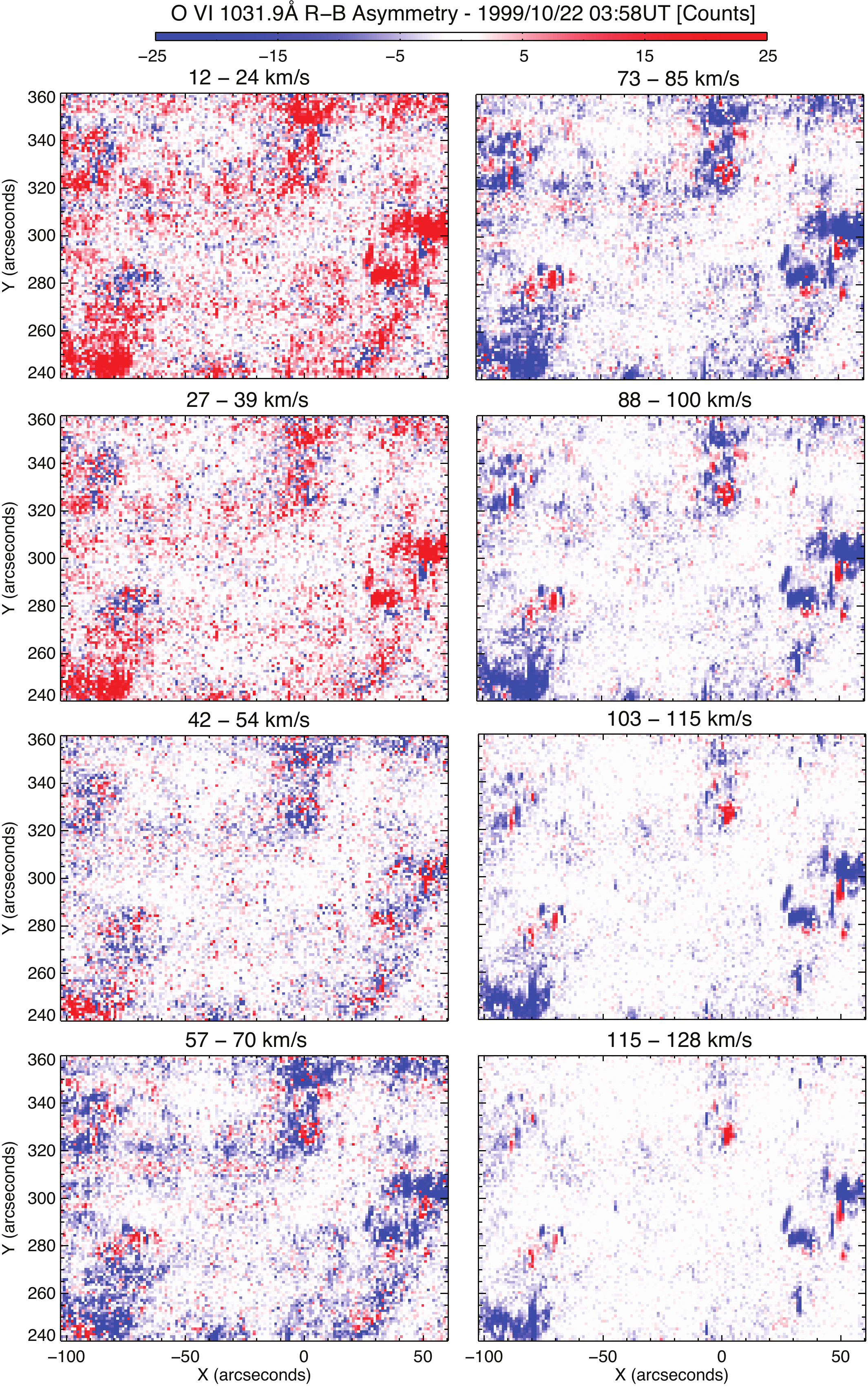}
\caption{Red-blue (R-B) profile asymmetries for the \ion{O}{6} 1031.9~\AA{} emission line at a range of velocities for the October 22 1999 (01:57-06:00UT) observation of an ECH. The electronic edition of the journal has movies showing the panels of this figure along with those at intermediate velocities and the components of Fig.~\pref{f15}. \label{f16}} 
\end{figure}

The observations presented in Figs.~\pref{f13_o6} through~\pref{f16} provide supporting evidence that the profile asymmetries in coronal holes have the same characteristic nature in ECH plasma in the \ion{O}{6} 1031.9~\AA{} line. The observations taken by SUMER of a small patch of ECH were taken on October 19-20 1999 (21:13-01:15UT) and October 22 1999 (01:57-06:00UT), the first to the East and the second to the West of the equator. The observations in both cases used Detector A, a $0.3\arcsec \times 120\arcsec$ slit, 1.13\arcsec\ horizontal steps, and 100s exposures. Figs.~\pref{f13_o6} and Figs.~\pref{f15} show context (supplied by EIT 195\AA{} images) and the result of a single Gaussian fit to the SUMER spectra. The location of the small rasters are shown as an orange box on the EIT image (left) image with the peak line intensity, (relative - not absolutely calibrated) Doppler velocity, and line width for that region shown top-to-bottom on the right. On inspection of the R-B analysis of the line profiles in the region (in Figs.~\pref{f14} and~\pref{f16}) we see little evidence that would allow us to draw a coronal hole boundary in the region, the reason why one is not drawn. The asymmetry of the profiles behaves as we have seen before: red at low velocities, but a more pronounced blue asymmetry going out to greater than 100 km/s. At such high velocities we expect that there may be some very high velocity explosive events \citep[][]{Innes1997} around some of the (presumed) network elements. This was noted for quiet sun spectra of the same line \citep[Sect. 2.3 of][]{McIntosh2009b}.

\begin{figure} 
\epsscale{1.2}
\plotone{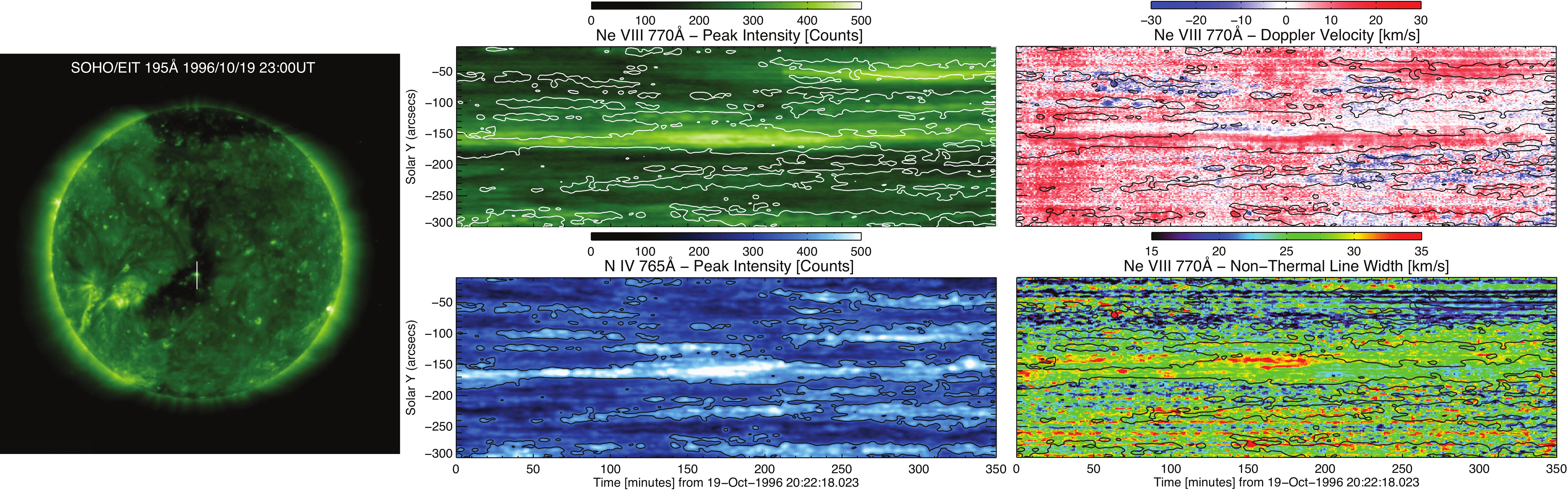}
\caption{The SUMER \ion{Ne}{8} 770~\AA{} ``drift scan'' of an equatorial coronal hole and EUV bright point in first spectral order observed between 20:22UT October 19 1996 and 03:24UT on October 20 1996. The left panel shows the EIT 195\AA{} image at the midpoint of the SUMER observation for context and the slit location is given by the orange vertical line. The panels on the right show the peak intensity of the \ion{Ne}{8} and \ion{N}{4} lines with the Doppler velocity, and the non-thermal line width of \ion{Ne}{8}. \label{f17}} 
\end{figure}

\begin{figure} 
\epsscale{1.2}
\plotone{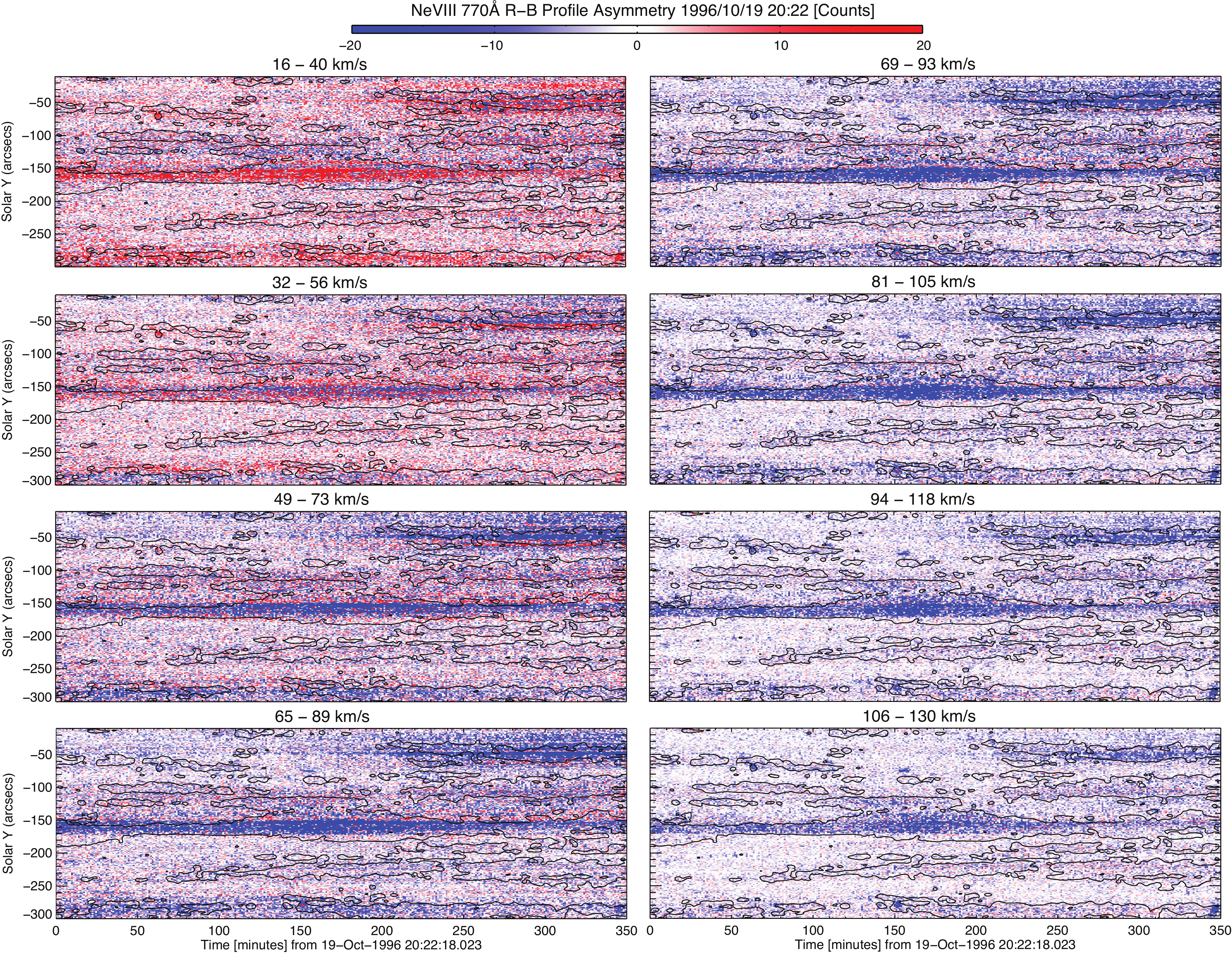}
\caption{Red-blue (R-B) profile asymmetries for the \ion{Ne}{8} 770.4~\AA{} emission line in first spectral order at a range of velocities. The contours shown in each panel are for the \ion{N}{4} 765\AA{} line at a iso-intensity level of 200. The electronic edition of the journal has movies showing the panels of this figure along with those at intermediate velocities and the components of Fig.~\pref{f17}.\label{f18}} 
\end{figure}

Figures~\pref{f17} through~\pref{f18} show the context and analysis of first order (Detector B) observations in an ECH (including an EUV bright point) in the \ion{Ne}{8} 770~\AA{} line on October 19-20 1996 (20:22 - 03:24UT) with an exposure time of 60s. Unfortunately, the malfunction in the scanning mechanism meant that this observation requires the Sun to rotate underneath the SUMER slit to ``map'' the region in a `drift-scan' \-- in this case the SUMER slit was held fixed at the central meridian and spectra were acquired allowing the Sun to rotate underneath and an image to be formed over time. We note that this observation includes supporting information from the \ion{N}{4} 765\AA{} line to provide some network context.  Figure~\pref{f17} shows the context EIT 195\AA{} from 23:00UT (left) with the slit position indicated. The panels on the right side of the figure show the \ion{Ne}{8} peak line intensity, Doppler velocity, non-thermal line width and the \ion{N}{4} peak intensity to isolate the network using an iso-intensity contour of 200. The passage of the EUV bright point and a few network elements inside the coronal hole are clear. 

Figure~\pref{f18} shows the R-B analysis of the first order \ion{Ne}{8} 770~\AA{} observation. We notice that the non-thermal line widths of the first order observation are consistent with those observed in the ECH in second order - indicating that the \ion{Si}{1} blends may not play a strong role in modifying the second order spectra. Similarly, we see a progression from red asymmetries to blue at higher speeds than for \ion{C}{4} in the coronal hole. More importantly, we find that many of the network elements (as outlined by \ion{N}{4} intensity contours) indeed show blueward asymmetries. This supports the results of the main body of the Paper. The timeseries also suggests that the blueward asymmetries are temporally intermittent, which we have noted before for quiet Sun \citep{McIntosh2009b}. The lack of blends in first order for this line permits observation of the extended blue wing of the line to higher velocities. Indeed, it seems as though the core of the EUV bright point (y = -160\arcsec) still shows a strong blue-wing asymmetry beyond 120km/s. This may indicate that the velocity range of the blue asymmetries depends on the magnetic field strength, which would suggest that the speed of the disturbances (e.g., spicules) in the lower atmosphere is dependent on how strong the field is. Alternatively, the larger velocities in the bright point may be because of a favorable viewing angle.

\end{document}